\documentclass[aps,prb,a4paper,twocolumn,10pt,superscriptaddress,notitlepage]{revtex4-2}

\newcommand{\papertitle}{Probing Azimuthal Anatomy of Hyperbolic Whispering Gallery Modes in hBN}

\usepackage[utf8]{inputenc}
\usepackage{graphicx}
\usepackage{wrapfig}
\usepackage{amsmath}
\usepackage{amssymb}
\usepackage{caption}
\usepackage{natbib}
\usepackage{float}
\usepackage{tikz}
\usepackage{stfloats}
\usepackage{siunitx}
\usepackage{dcolumn}
\usepackage{lmodern}
\usepackage[T1]{fontenc}
\usepackage[unicode=true,
	    colorlinks=true,
	    linkcolor=blue,
	    citecolor=blue,
	    urlcolor=blue]{hyperref} 
\usepackage{sansmath}
\usepackage{blindtext}
\usepackage[normalem]{ulem}

\usepackage{chemformula}
\usepackage{empheq}
\usepackage{comment}
\usepackage{bm}
\usepackage{gensymb}

\usepackage{caption}
\usepackage[misc]{ifsym}
\usepackage{fontawesome}

\usepackage{lineno}

\setcitestyle{super}

\usepackage{xr}
\makeatletter

\newcommand*{\addFileDependency}[1]{
\typeout{(#1)}
\@addtofilelist{#1}
\IfFileExists{#1}{}{\typeout{No file #1.}}
}\makeatother

\newcommand*{\myexternaldocument}[1]{%
\externaldocument{#1}%
\addFileDependency{#1.tex}%
\addFileDependency{#1.aux}%
}

\myexternaldocument{supplement}

\usepackage[top=2.5cm,bottom=2.5cm,left=1.5cm,right=1.5cm]{geometry}

\begin{document}

\title{\Large\textsf{\papertitle}}

\author{Bogdan Borodin}
\email{bborodin@nd.edu}
\affiliation{\footnotesize Department of Physics and Astronomy, University of Notre Dame, Notre Dame, IN 46556,~USA}
\affiliation{\footnotesize Stavropoulos Center for Complex Quantum Matter, University of Notre Dame, Notre Dame, IN 46556,~USA}

\author{Samyobrata Mukherjee}
\affiliation{\footnotesize School of Applied and Engineering Physics, Cornell University, Ithaca, NY 14853, USA}

\author{Shivaksh Rawat}
\affiliation{\footnotesize School of Applied and Engineering Physics, Cornell University, Ithaca, NY 14853, USA}

\author{Seojoo Lee}
\affiliation{\footnotesize School of Applied and Engineering Physics, Cornell University, Ithaca, NY 14853, USA}
\affiliation{\footnotesize The Institute of Basic Science, Korea University, Seoul 02841, Republic of Korea}

\author{Thomas Poirier}
\affiliation{\footnotesize  Tim Taylor Department of Chemical Engineering, Kansas State University, Manhattan, 66506,
Kansas, ~USA}

\author{Kenji Watanabe}
\affiliation{\footnotesize Research Center for Electronic and Optical Materials, National Institute for Materials Science, 1-1 Namiki, Tsukuba 305-0044,~Japan}

\author{Takashi Taniguchi}
\affiliation{\footnotesize Research Center for Materials Nanoarchitectonics, National Institute for Materials Science,  1-1 Namiki, Tsukuba 305-0044,~Japan}

\author{James H. Edgar}
\affiliation{\footnotesize  Tim Taylor Department of Chemical Engineering, Kansas State University, Manhattan, 66506,
Kansas, ~USA}

\author{Hanan Herzig Sheinfux}
\affiliation{\footnotesize  Department of Physics, Bar-Ilan University, Ramat Gan 52900, Israel}

\author{Gennady Shvets}
\affiliation{\footnotesize School of Applied and Engineering Physics, Cornell University, Ithaca, NY 14853, USA}

\author{Petr Stepanov}
\email{pstepano@nd.edu}
\affiliation{\footnotesize Department of Physics and Astronomy, University of Notre Dame, Notre Dame, IN 46556,~USA}
\affiliation{\footnotesize Stavropoulos Center for Complex Quantum Matter, University of Notre Dame, Notre Dame, IN 46556,~USA}



\begin{abstract}
\noindent
Scattering-type scanning near-field optical microscopy (s-SNOM) is a powerful tool for investigating polaritonic modes. However, an inherent limitation of this technique is that excitation and detection occur at the same location. This constraint makes it challenging to resolve excitations with complex spatial structures, which can exhibit delicate dependence on the in-coupling conditions. Here, we present a strategy to overcome this limitation by introducing an auxiliary cavity, which serves as a stationary near-field excitation source. This configuration introduces a tip-independent and spatially separated excitation source, allowing the s-SNOM tip to primarily probe the resulting modal field distribution with minimal perturbation of the resonator. We apply this approach to whispering gallery modes (WGMs) of hyperbolic phonon-polaritons in hexagonal boron nitride resonators. Through spatially resolved near-field maps, we directly observe subwavelength polaritonic WGMs with large and discrete azimuthal momentum ($k_\phi / k_0$ up to 15). Direct access to the modal field distributions enables identification of the azimuthal mode numbers and their frequency-dependent excitation spectra. We further show that, over finite spectral intervals, the modes continuously adjust their effective refractive index while preserving a fixed azimuthal mode number before transitioning to the next resonant state. Numerical simulations support the experimental observations and confirm the observation of hyperbolic WGMs. This approach enables direct visualization of previously hidden mode structures in hyperbolic cavities and opens new opportunities for studying and engineering confined polaritonic states.
\end{abstract}


\maketitle
\vspace{-2em}
\section{Introduction}
\vspace{-1em}
Materials whose permittivity has opposite signs along different principal axes are known as hyperbolic materials. These materials support hyperbolic excitations with extremely large momenta. In particular, in the vicinity of phonon resonances these polaritons can simultaneously exhibit large momenta and relatively low losses\cite{dai2014tunable,jacob2014hyperbolic}, making hyperbolic phonon polaritons (HPhPs) particularly attractive for mid-infrared and terahertz nanophotonics. This includes applications such as sub-diffraction-limited imaging, sensing, anisotropy engineering, and lensing\cite{huang2023plane, obst2023terahertz, feres2021sub,ni2021long,dai2018hyperbolic,dai2015subdiffractional, borodin2026cavity}.

Nanocavities can further enhance the desirable properties of polaritonic materials, and HPhPs are no exception. For example, multimodal HPhP nanocavities with quality factors (Q-factors) of up to 400, spin–orbit-locked hyperbolic polariton vortices\cite{herzig2024high,wang2022spin,conrads2024direct}, enabling topological charge transfer\cite{orsini2024deep}, programmable polaritonic cavities in phase-change materials\cite{chaudhary2019polariton}, and high-Q hyperbolic and surface phonon-polariton resonators in hBN ribbons\cite{dolado2020nanoscale} have been demonstrated. These studies highlight the remarkable capabilities of confined hyperbolic phonon-polaritons and establish nanocavities as a versatile platform for manipulating light at the nanoscale. However, the majority of these resonators are primarily based on low-order mode confinement. In contrast, in macroscopic optical resonators, higher-order whispering gallery modes (WGMs) are known for their exceptionally high quality factors. State-of-the-art structures can achieve Q-factors as high as $10^{11}$\cite{grudinin2006ultra,lin2014barium,savchenkov2007optical}. Such extraordinary performance has made WGMs instrumental for a wide range of applications in both fundamental and applied photonics. Recently, hyperbolic WGMs have also been observed spectroscopically in hBN nanotubes\cite{guo2023hyperbolic}, demonstrating that this concept can be extended to hyperbolic polaritons. However, direct experimental access to the spatial field distributions of individual hyperbolic WGMs has remained elusive.
Combining these concepts offers the possibility of hyperbolic nano-WGMs that support extraordinarily large and discrete azimuthal momentum \(k_\phi\) while maintaining relatively high Q-factors. Unlike propagating hyperbolic phonon-polaritons, WGMs form a discrete set of cavity eigenstates characterized by quantized azimuthal momentum. Such confined modes provide an attractive platform for studying high-\(k\) polaritonic states, strong light confinement, and resonant light--matter interactions at the nanoscale. Their discrete nature also offers a clear framework for investigating near-field interference, mode hybridization, and coupling between polaritonic nanostructures. Direct visualization of these field distributions requires a launcher capable of efficiently exciting the resonator without significantly perturbing its modal structure, together with a characterization technique providing deep sub-diffraction spatial resolution. Such capability would enable direct identification of the mode order, spatial field distribution, and excitation characteristics of hyperbolic WGMs.

s-SNOM is a powerful tool for probing the spatial distribution of sub-diffraction polaritonic modes, yet it suffers from a fundamental limitation. The metallic tip simultaneously launches and probes phonon-polaritons, making the measured near-field signal inherently dependent on the excitation conditions. Several remote-excitation strategies have been developed to alleviate this limitation, including metallic antennas, buried metallic structures, wrinkles, and other engineered launchers\cite{pons2019launching,duan2017launching,ni2021long}. These approaches have proven highly successful for launching propagating hyperbolic phonon-polaritons. However, their application to isolated resonant nanocavities remains challenging, as many launchers either directly perturb the resonator or remain connected to it through the surrounding hBN, providing leakage pathways that can modify the cavity response. As recently shown\cite{herzig2023transverse}, a periodic metallic structure on hBN forms a lattice that profoundly modifies the mode profile and momentum. Consequently, obtaining direct experimental access to the intrinsic azimuthal field distribution of hyperbolic WGMs remains challenging. Overcoming this limitation would establish a reliable approach for studying, manipulating, and engineering whispering-gallery modes in polaritonic resonators.

In this work, we present an auxiliary-cavity-assisted excitation strategy based on a resonant cavity formed at a metal--dielectric interface (Au/SiO$_2$). The auxiliary cavity serves as a stationary near-field excitation source that is independent of the tip position. Thus, the auxiliary cavity resonantly enhances the local near field and launches propagating hyperbolic phonon-polaritons whose angular distribution is governed by the cavity geometry. This enables efficient coupling to whispering-gallery modes without significantly modifying their intrinsic resonance conditions.

Consequently, we obtain spatially resolved near-field maps that reveal hyperbolic high-Q WGMs with large and discrete azimuthal momentum ($k_\phi / k_0$ up to 15) in subwavelength cavities. Direct visualization of the modal field distributions enables unambiguous identification of the azimuthal mode numbers and their frequency-dependent excitation spectra. WGMs with large azimuthal numbers are shown to be tightly confined within the resonators. Fourier-transform analysis reveals the frequency-dependent excitation spectra and evolution of the excited WGM states. By tuning the excitation frequency, we switch between WGMs with different azimuthal and radial numbers. Moreover, within the linewidth of a given mode, the azimuthal mode number remains unchanged while the effective refractive index evolves continuously before the resonator transitions to the neighboring whispering-gallery mode.

\vspace{-2em}
\section{Results}
\vspace{-1em}
\subsection{Principle of the tip-independent excitation}
\vspace{-1em}

The concept of tip-independent excitation and detection is illustrated in Fig.~1. We compare this method with the conventional observation of radial polaritonic interference fringes in s-SNOM measurements. Such radial fringes are most prominent near sample edges, exhibit a \( \lambda/2 \) periodicity, and correspond to so-called tip-launched round-trip (RT) modes. The measured signal originates from the interference between the polariton launched by the tip and the same polariton after reflection from the sample edge (see the inset in Fig.~1a). The interference condition is given by $\Delta L \cdot n_{\mathrm{eff}} = N\cdot\lambda/2$, where \( \Delta L \) is the optical path difference, \( N \) is an integer, \( n_{\mathrm{eff}} \) is the effective refractive index, and \( \lambda \) is the polariton wavelength. As the tip is displaced by \( \lambda/2 \), the total optical path length changes by \( \lambda \) (accounting for both the forward and reflected paths), resulting in a \( \lambda/2 \)-periodic interference pattern (see the inset \textit{SNOM Amplitude} in Fig.~1a).
\onecolumngrid
\vspace{+2em}
\begin{center}
\includegraphics[width=1.0\textwidth]{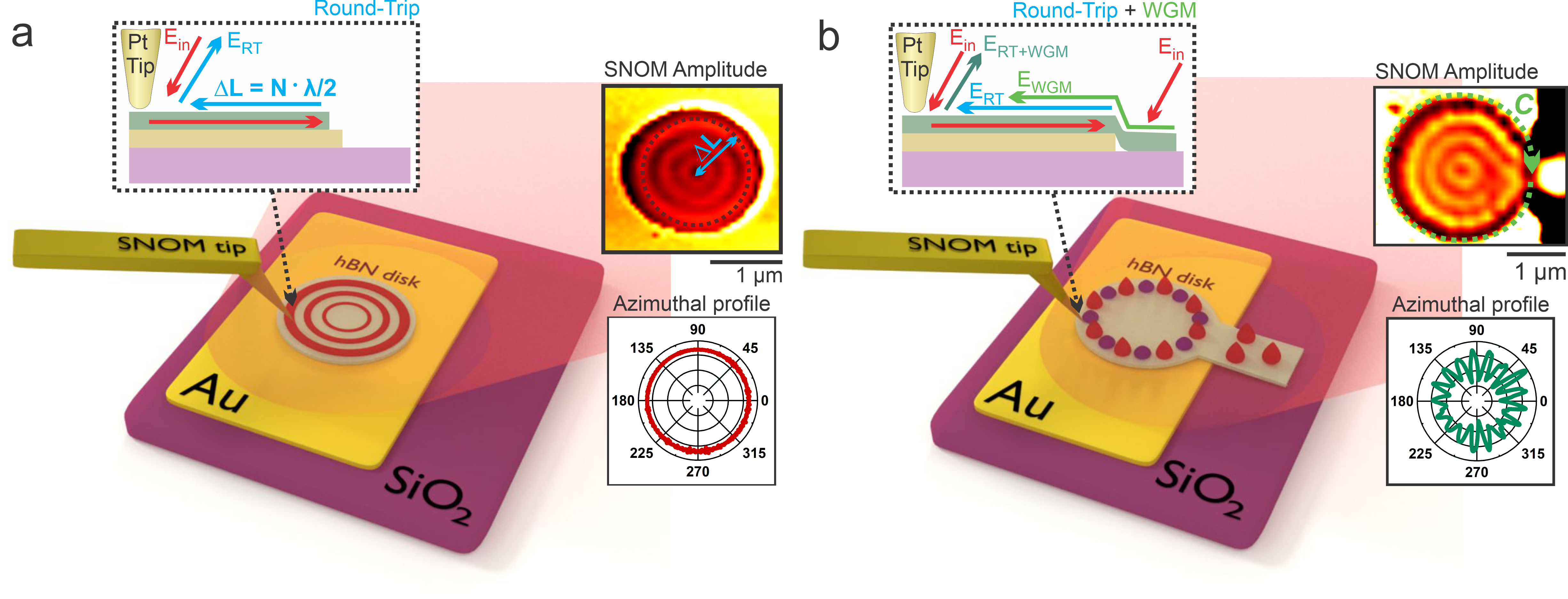} \\[1ex]
\parbox{1\textwidth}{\setlength{\parindent}{0pt}%
\textbf{Figure 1: Principle scheme of the s-SNOM experiment.}
\textbf{a}) Standard s-SNOM for a standalone hBN disk. Tip-launched round-trip (RT) modes produce radial intensity modulation. Insets show the experimental s-SNOM amplitude map and the corresponding azimuthal profile near the disk rim.
\textbf{b}) s-SNOM for an hBN disk with an auxiliary cavity providing tip-independent excitation. Along with the RT mode, a cavity-launched WGM independent of the tip position produces azimuthal modulation. Insets show the experimental s-SNOM amplitude map and the corresponding azimuthal profile near the disk rim.
}
\end{center}
\twocolumngrid
\noindent
This Fabry-Pérot-like round-trip resonance has been observed in a wide variety of systems, ranging from one-dimensional nanotubes to physically defined cavities and cavities formed by dielectric contrast\cite{phillips2021fabry,ma2024plane,herzig2024high,conrads2024direct,jackering2025tailoring}. In hBN disks, this type of resonance has been widely used to study the influence of anisotropic substrates on hyperbolic phonon-polaritons\cite{chaudhary2019engineering}. An alternative approach relies on remote excitation of polaritons using engineered launchers, including metallic structures, buried metallic disks, wrinkles, and plasmonic antennas\cite{pons2019launching,duan2017launching,ni2021long,huber2008focusing}. In these configurations, the measured signal arises from the interference between the polariton launched by the external feature and detected by the s-SNOM tip, and the time-reversed path, in which the polariton is launched by the tip and scattered by the launcher. Consequently, the measured fringes exhibit periodicity \( \Delta L \). The influence of the launcher on the resonator depends on the specific excitation geometry. While several remote-excitation approaches have successfully demonstrated efficient launching of propagating hyperbolic phonon-polaritons, their application to isolated resonant nanocavities remains challenging because many launchers either directly perturb the resonator or remain connected to it through the surrounding hBN, providing leakage pathways that can modify the cavity response. In contrast, the auxiliary cavity employed in this work is separated from the whispering-gallery resonator by a metal/dielectric impedance-mismatch interface and therefore serves as a resonantly enhanced near-field launcher while leaving the resonator itself free of perturbing excitation structures. We further investigate the role of the impedance-mismatched interface and the auxiliary cavity geometry in Section I of the Supplementary Information.

In circular WGM resonators, polaritons are localized near the disk boundary, and the resonance condition can be written as
$C \cdot n_{\mathrm{eff}} = m\cdot\lambda$, where \(C\) is the circumference. Constructive interference occurs when an integer number of wavelengths fits within this optical path, leading to the formation of a standing wave with azimuthal field modulation near the rim. Thus, the well-known smoking-gun signatures of WGMs are: (i) azimuthal field modulation, (ii) strong localization at the cavity edge, (iii) quasi-equidistant mode switching, and (iv) \(m\) is approximately given by \(m \approx n_{\mathrm{eff}}k_0R\)\cite{wait1967electromagnetic}. To overcome the above-discussed limitations, we adopt an alternative strategy\cite{borodin2026cavity}.

By fabricating an auxiliary cavity at the Au/SiO$_2$ interface, we introduce a tip-independent resonant near-field excitation source. The cavity supports a localized resonant mode that efficiently couples to the far field. Because the Au/SiO$_2$ interface is only partially reflecting, this resonant mode is not completely confined and launches propagating hyperbolic phonon-polaritons into the hBN layer. Rather than defining the allowed WGM momenta, the auxiliary cavity enhances the excitation efficiency of the resonant modes supported by the disk. The total near-field inside the hBN disk can then be written as $E_{\mathrm{total}}=E_{\mathrm{tip}}+E_{\mathrm{cav}}$ where \(E_{\mathrm{tip}}\) corresponds to the conventional tip-launched RT mode, while \(E_{\mathrm{cav}}\) originates from the stationary auxiliary cavity and is independent of the tip position. The two excitation channels coexist, allowing the WGM field distribution to be observed without relying solely on tip-launched polaritons (see the upper-left inset in Fig.~1b).

This results in a field distribution containing both the RT mode originating from the tip-scattering process and an additional field modulation near the cavity edge. The RT and WGM contributions coexist and interfere, while the latter is localized near the resonator boundary. The modulation near the edge results from the interference between the WGM launched by the auxiliary cavity (see the inset \textit{SNOM Amplitude} in Fig.~1b) and its time-reciprocal path. The inset \textit{Azimuthal profile} in Fig.~1b shows the amplitude profile taken along the circumference in the vicinity of the disk edge. As can be seen, this profile reveals clear azimuthal field modulation characteristic of WGMs.

\vspace{-1.5em}
\subsection{Azimuthal field distribution analysis}
\vspace{-1em}

Figure~2a shows the topography of a 1-\(\mu\)m-radius, 32 nm thick hBN disk positioned on top of a gold stripe, which is connected to a rectangular hBN cavity resting on the SiO\(_2\) substrate. Figure~2b presents the s-SNOM optical amplitude demodulated at the third harmonic of the tip oscillation (\(S_3\)), which is proportional to the out-of-plane near-field intensity, \( |E_z|^2 \), of the structure. As was discussed above, besides the concentric circles originating from the RT mode, the map shows periodic azimuthal modulation along the rim. The inset displays a 3D surface plot to highlight this periodic modulation more clearly. This modulation arises from hyperbolic phonon-polariton whispering-gallery modes launched by the rectangular auxiliary cavity.

To analyze the periodicity and characterize the mode, we extract azimuthal profiles along multiple circumferences spanning the entire resonator, from the cavity edge toward the center (see Fig. 2c). Each circumference is analyzed independently, allowing us to follow the evolution of the WGM signal as a function of radial position.

The measured near-field signal contains contributions from both the conventional RT mode and the WGM. The RT mode, together with the finite intensity variation introduced by the auxiliary cavity, produces a slowly varying background that changes from one circumference to another, whereas the WGM appears as a periodic azimuthal modulation. To compare the modulation strength at different radii, we subtract a smooth polynomial (third order) baseline from each azimuthal profile. This procedure removes only the slowly varying background (the zero-spatial-frequency component) while preserving the finite-spatial-frequency modulation associated with the WGMs.  Figure~2c demonstrates the profiles extracted after the baseline subtraction. As can be seen, the green profiles extracted at R $\approx$ 0.9 $\mu$m show the strongest azimuthal modulation, even though their absolute near-field intensity is relatively low, whereas the modulation gradually decreases for profiles extracted closer to the cavity center.
\onecolumngrid
\vspace{-1.5em}
\noindent
\begin{center}
\includegraphics[width=0.97\textwidth]{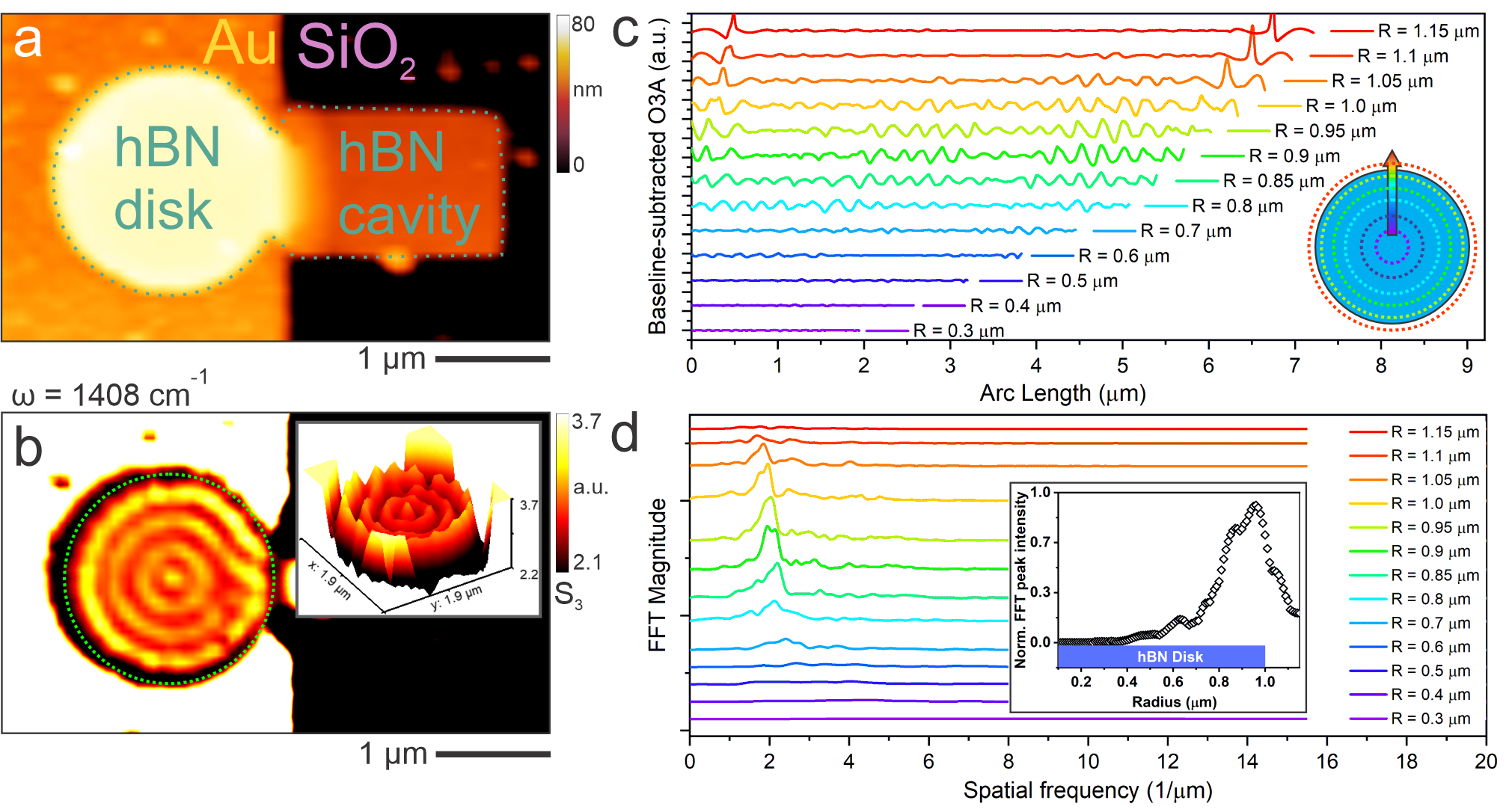}\\[0ex]
\parbox{1\textwidth}{\setlength{\parindent}{0pt}%
\textbf{Figure 2: Detection and analysis of HPhP whispering-gallery modes.}
\textbf{a}) AFM topography of a 1-\(\mu\)m-radius, 32-nm-thick hBN disk positioned on top of a 50-nm-thick gold stripe connected to a rectangular hBN cavity resting on the SiO\(_2\) substrate.
\textbf{b}) s-SNOM optical amplitude at 1408 cm\(^{-1}\) demodulated at the third harmonic of the tip oscillation (\(S_3\)), which is proportional to the out-of-plane near-field intensity, \( |E_z|^2 \), in the structure. The dashed circle marks the radial position \(R\approx0.9~\mu\mathrm{m}\), where the azimuthal WGM modulation is strongest. The inset shows a 3D representation of panel b.
\textbf{c}) Azimuthal profiles extracted along circumferences schematically shown in the inset after smooth baseline removal. The complete set of analyzed profiles is shown, illustrating the evolution of the WGM modulation from the cavity edge toward the center.
\textbf{d}) Fourier spectra of the baseline-subtracted profiles. The inset shows the FFT peak amplitude as a function of the circumference radius (bandwidth $\delta$R = 0.1 um, r$_{step}$ = 0.01 um).
}
\end{center}
\twocolumngrid
\noindent
Figure~2d shows the Fourier spectra corresponding to these profiles. As expected, the FFT peak quenches for profiles taken closer to the center. For each analyzed circumference, the WGM-related FFT peak is identified independently. The corresponding azimuthal wavevector is determined from the peak position according to $k_{\phi}=m/r$, where $m$ is the azimuthal mode number and $r$ is the radius of the analyzed circumference. The inset in Fig.~2d summarizes the evolution of the normalized WGM-related FFT peak amplitude as a function of the circumference radius. Consistently with a WGM, we observe that the FFT peak is strongly localized near the cavity edge. It is important to note that the FFT amplitude reaches its maximum at the geometrical edge of the resonator, consistent with the expected localization of whispering-gallery modes. Within the cavity, the radial dependence of the FFT peak directly reflects the spatial distribution of the WGM, with the strongest modulation occurring near the sidewall where the mode is confined. At the same time, a finite FFT signal remains detectable slightly outside the resonator. This does not indicate that the WGM propagates outside the disk, but rather reflects the evanescent field extending beyond the cavity boundary. As confirmed by the effective-index modeling (see Figure S11 in Supplementary Information), the modal field penetrates into the surrounding medium, allowing the near-field probe to efficiently couple to the circulating mode when positioned close to the cavity sidewall. Such evanescent-wave coupling is well known for optical whispering-gallery resonators and forms the basis of their efficient excitation and readout~\cite{lin2017nonlinear,vollmer2008whispering}.

Typically, WGM resonances are described by two integers, \(m\) and \(n\), the azimuthal and radial mode numbers, respectively. The azimuthal number \(m\) specifies the number of optical wavelengths (or phase cycles) along the azimuthal direction around the resonator's edge. The corresponding azimuthal wavevector is given by \(k_{\phi}=m/r\), where \(r\) is the radius of the analyzed circumference. The radial number \(n\) counts the number of radial field maxima; higher-\(n\) modes have their energy concentrated at a smaller effective radius, so they circulate along a shorter optical path and are less tightly confined to the boundary. As a result, modes with large \(m\) and low \(n\) exhibit the strongest confinement and typically achieve the highest Q-factors.

It is important to emphasize that, in pseudo-heterodyne s-SNOM, the detected complex near-field signal is reconstructed from the in-phase and quadrature components, \(S=I+iQ\), with amplitude \(A=\sqrt{I^{2}+Q^{2}}\) and phase \(\phi=\tan^{-1}(Q/I)\). Both amplitude and phase images are presented in Section II of the Supplementary Information. The phase images are presented later in Section C because they generally provide higher visual contrast of the whispering-gallery modes by suppressing slowly varying background contributions, making the modal field distribution easier to identify. However, the displayed phase represents the phase of the complex near-field signal relative to the interferometric reference rather than the algebraic sign of the optical field.

For the quantitative analysis, the FFT is performed on the near-field amplitude rather than the phase. The experimental uncertainties give \(\delta A=(I\,\delta I+Q\,\delta Q)/A\), whereas \(\delta\phi=(I\,\delta Q-Q\,\delta I)/(I^{2}+Q^{2})\propto1/A\). Consequently, the phase uncertainty increases rapidly in regions of weak near-field signal, while the amplitude remains comparatively robust, making it a more statistically reliable quantity for determining the modal periodicity. Since the amplitude represents the magnitude of the complex optical field, the positive and negative extrema of the underlying standing-wave field give rise to identical amplitude maxima. As a result, each optical cycle produces two maxima in the measured amplitude profile, and the dominant FFT peak corresponds to twice the physical azimuthal mode number. Throughout this work, the WGM mode number is therefore obtained as \(m=f_{\mathrm{FFT}}r/2\), where \(f_{\mathrm{FFT}}\) is the azimuthal spatial frequency extracted from the FFT and \(r\) is the resonator radius. Manual counting of the amplitude maxima yields identical mode assignments and serves as an independent verification of the FFT analysis. This approach enables direct experimental determination of the WGM quantum numbers, which previously were primarily inferred from numerical modeling of the resonant field distributions~\cite{ye2015monolayer,borodin2023indirect,eliseyev2023twisted,alekseev2025engineering}.

\vspace{-2em}
\subsection{Mode evolution and Q-factor of hyperbolic phonon-polariton WGMs}
\vspace{-1em}

Figure~3a shows representative s-SNOM optical phase maps demodulated at the fourth harmonic ($P_4$) for several excitation wavelengths. The corresponding amplitude maps are used for the quantitative FFT analysis discussed below. As the wavenumber $\omega$ increases, more concentric circles (RT mode) appear in the disk because the wavelength inside the material shrinks and constructive interference occurs over the same optical path. In addition, more peaks form along the disk's circumference, indicating an increase in the azimuthal mode number of the WGMs. The measured field distribution closely resembles the 3D driven simulations of the \( |E_z|^2 \) field distribution in the disk--cavity structure (see Fig.~S1 in the Supplementary Information). The simulations successfully reproduce both the round-trip mode and the azimuthal modulation observed experimentally. Driven simulations allow us to observe the full diversity of modes that can be excited in the structure as a whole, instead of studying specific eigenmodes. While this approach provides a mode distribution that closely matches the experiment, it does not allow us to reliably extract the azimuthal mode numbers \( m \) from the simulations and directly compare them with experiment. For this purpose, 2D eigenmode simulations can be used (see Numerical Simulations for details).

The analysis described in Section B was applied to s-SNOM amplitude maps recorded across excitation wavenumbers from 1370 to 1430 cm$^{-1}$, with a step size of 2 cm$^{-1}$. The resulting non-normalized FFT spectra were used to generate the contour plot shown in Fig.~3b. Each trace represents an individual FFT spectrum, where color encodes the magnitude of the Fourier peak at a given wavenumber and wavevector. Thus, the intensity of peaks in Fig.~3b reflects the strength of the azimuthal field modulation within the cavity.

Rather than a direct spectroscopic measurement, the raspberry curve shown in Fig.~3c is constructed from the FFT analysis presented in Fig.~3b. For each excitation wavenumber, the FFT spectrum is analyzed over the full range of detected azimuthal momenta, and the dominant WGM peak is identified and its amplitude extracted, yielding one experimental point per excitation wavelength. Figure~3c therefore tracks the amplitude of the dominant WGM mode as the excitation wavenumber is varied. The resulting dependence represents a pseudo-spectroscopic response, in which the resonance strength is reconstructed from the spatial near-field modulation rather than measured directly by conventional far-field spectroscopy. Such reconstruction of spectroscopic information from near-field imaging is widely used in s-SNOM studies of polaritons owing to the imaging nature of the technique~\cite{herzig2024high,wang2020revealing,chaudhary2019engineering}. The raspberry-colored diamonds correspond to the experimentally extracted FFT amplitudes, while the solid curve represents a fit using a sum of quasi-equidistant Lorentzian resonances. The quasi-equidistant peaks originate from successive whispering-gallery modes supported by the disk resonator. Because these resonances partially overlap spectrally, Lorentzian decomposition is used to separate the individual modal contributions and determine their resonance frequencies and linewidths. The quality factors are obtained from the fitted Lorentzian linewidths according to \(Q=f_0/\Delta f\). As discussed in Section III of the Supplementary Information, model-selection analysis based on the Bayesian Information Criterion confirms that the chosen Lorentzian decomposition provides the optimal balance between fit quality and model complexity. Notably, Q-factors exceed 300 for several modes, among the highest reported for HPhP resonances.

Figure~3d shows normalized FFT spectra, identifying the dominant mode at each excitation wavenumber, independent of variations in the absolute amplitude of the FFT peaks. This representation is particularly useful for tracking the most pronounced resonances without the influence of intensity fluctuations. Raspberry-colored diamonds indicate the exact positions of the FFT peaks. Finally, the azimuthal mode numbers for the dominant resonances were determined using the FFT procedure described in Section~B, with manual counting of the real-space field maxima serving as independent verification. The \((m,n)\) mode numbers are shown in brackets.

As observed, in the range from 1370~cm$^{-1}$ to approximately 1400~cm$^{-1}$, only a single dominant peak is visible for each excitation wavenumber. Above $\sim$1400~cm$^{-1}$, multiple distinct peaks emerge, reflecting the simultaneous excitation of several whispering-gallery modes with different mode indices at shorter wavelengths. The presence of multiple resonances explains the reduced FFT amplitude in Fig.~3b, as the field becomes distributed among several competing modes with different azimuthal wavevectors.
\onecolumngrid
\noindent
\begin{center}
\includegraphics[width=0.98\textwidth]{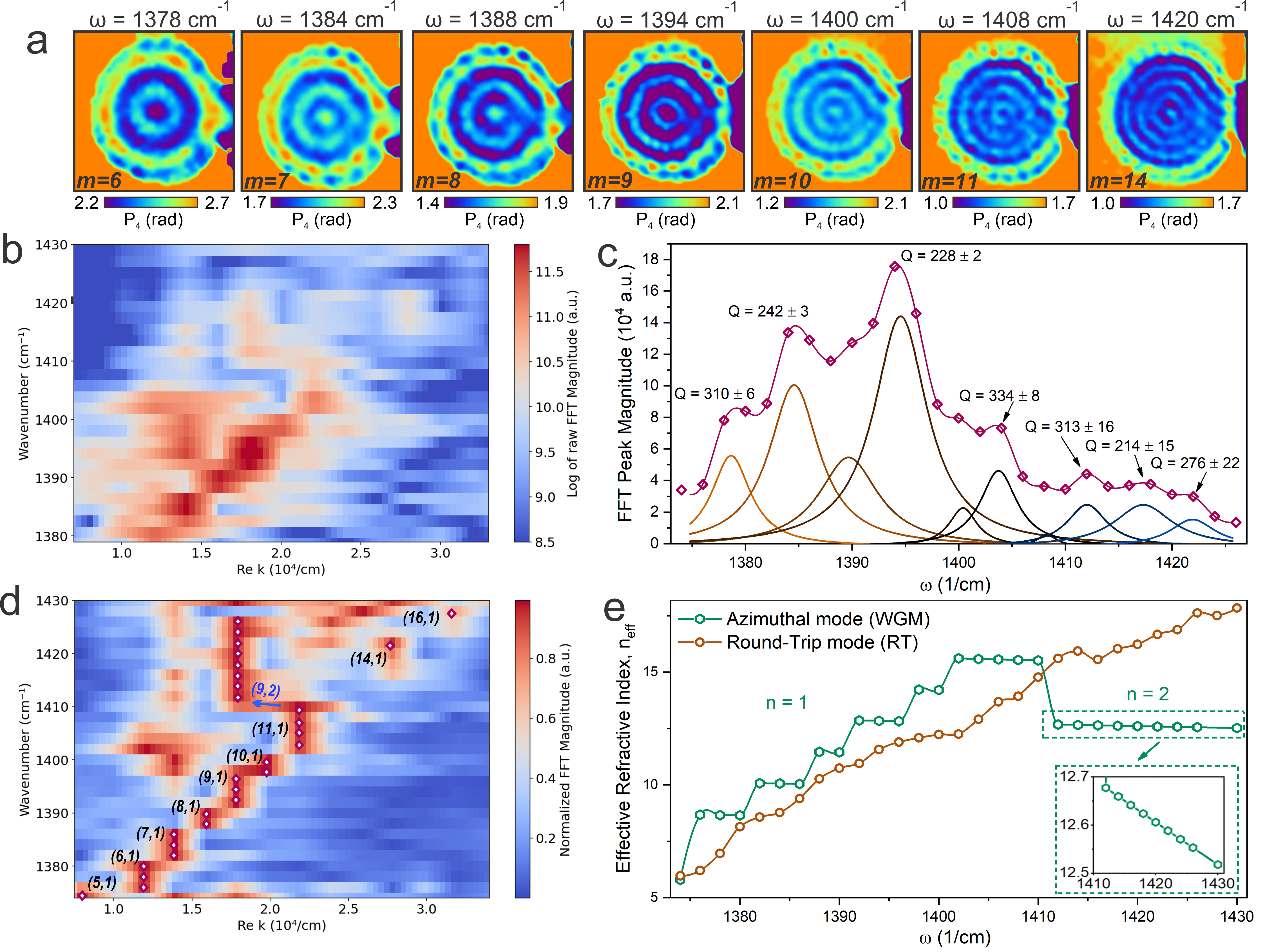} \\[1ex]
\parbox{1\textwidth}{\setlength{\parindent}{0pt}%
\textbf{Figure 3: Pseudo-spectroscopic analysis and mode evolution of HPhP WGMs.}
\textbf{a}) Representative s-SNOM optical phase maps demodulated at the 4$^{\text{th}}$ harmonic ($P_4$) for excitation wavelengths corresponding to modes with distinct $m$. The corresponding amplitude maps were used for the FFT analysis.
\textbf{b)} Contour plot constructed from raw FFT spectra (magnitude on a logarithmic scale) of profiles taken along the disk circumference, recorded over excitation wavenumbers from 1370 to 1430~cm$^{-1}$ with a step size of 2~cm$^{-1}$.
\textbf{c}) Reconstructed pseudo-spectroscopic response obtained from the FFT analysis. The raspberry-colored diamonds represent the amplitude of the dominant WGM mode extracted at each excitation wavenumber, while the solid curve represents a sum of Lorentzian resonances. Q-factors were estimated as the ratio of peak position to full width at half maximum ($f_0/\delta f$).
\textbf{d}) Normalized FFT spectra. Raspberry-colored diamonds indicate the exact positions of FFT peaks. The mode numbers (determined from the FFT analysis and verified by manual counting) are shown in brackets.
\textbf{e}) Effective refractive indices for azimuthal (WGM) and Round-Trip (RT) resonances. The inset shows dynamic refractive index tuning for the mode (9,2).
}
\end{center}
\twocolumngrid
\noindent
Despite the increased modal complexity, the dominant resonances remain clearly identifiable in Fig.~3d. Rather than representing the continuous dispersion of propagating HPhPs, the step-like dependence reflects the frequency-dependent excitation of discrete cavity eigenmodes. Within each plateau, the same whispering-gallery mode remains the dominant excitation channel over a finite spectral interval. As the excitation frequency is tuned further, coupling to this mode gradually becomes less favorable until another mode exhibits lower overall losses and becomes the dominant resonance, producing the abrupt transition to a new plateau. Thus, Fig.~3d represents the evolution of the most efficiently excited cavity mode as a function of excitation frequency. Within each plateau, the field distribution remains nearly unchanged while the effective refractive index gradually adjusts to maintain the resonance condition. Such continuous tuning of the effective modal index within the linewidth of a localized resonance is a general property of resonant cavities and reflects the gradual redistribution of the modal field between the resonator and its evanescent tail. Similar behavior has recently been reported for multimode hyperbolic nanocavities, where the same standing-wave pattern persists over a finite spectral interval before abruptly switching to the next mode order~\cite{herzig2024high}.

It should also be noted that after $\sim1410\,\mathrm{cm}^{-1}$ the dominant resonance exhibits a smaller number of azimuthal maxima than expected from the monotonic increase of $m$. Therefore, we attribute this behavior to competition between the first and second radial mode families. In this frequency range, both the $(9,2)$ and $(14,1)$ resonances are observed. Switching between radial mode families is a characteristic feature of whispering-gallery resonators~\cite{borodin2023indirect,alekseev2025engineering,vollmer2008whispering}.

Figure~3e displays the effective refractive indices of the WGM and round-trip modes extracted directly from the experiment. As can be seen, the round-trip modes exhibit a near-linear dependence in the presented region, which is in good agreement with the normal HPhP dispersion in hBN. The round-trip modes thus provide a reference for the underlying propagating HPhP dispersion in this spectral range. Meanwhile, the azimuthal dispersion consists of a series of plateaus near the HPhP dispersion line. Within each plateau, the number of azimuthal field maxima remains constant, while the effective refractive index is gradually tuned (see inset in Fig.~3e). As the excitation wavelength decreases, the effective index also decreases, maintaining a fixed number of optical cycles around the circumference. This behavior is consistent with the WGM resonance condition
\(m\lambda = 2\pi R n_{\mathrm{eff}}\), in agreement with the mode evolution discussed above. Here, $n_{\mathrm{eff}}$ describes the effective refractive index of the confined polaritonic mode, rather than the material refractive index of hBN. Longer plateaus are generally observed for higher-$m$ modes because a smaller relative variation of the effective refractive index is required to satisfy the resonance condition, i.e.,
\(\Delta\lambda/\lambda \propto \Delta n_{\mathrm{eff}}/n_{\mathrm{eff}}\). The experimentally observed $(9,2)$ mode, however, remains stable over an unusually broad spectral interval. We attribute this behavior to its weaker localization at the etched cavity boundary compared with the competing high-$m$ first-radial-order modes. Since higher-$m$ modes are more tightly confined to the etched sidewall, they are expected to experience stronger scattering from edge imperfections, allowing the second-radial-order mode to remain favorable over a broader spectral range. Consequently, the $(9,2)$ mode exhibits a characteristic non-monotonic shift toward lower \(n_{\mathrm{eff}}\), reflecting the transition from the $(m,1)$ to the $(m,2)$ radial mode family. The positive dispersion of the round-trip modes in the same spectral range confirms the conventional positive HPhP group velocity, distinguishing it from the negative slope of the WGM effective index.

The mode with $n=2$ is localized closer to the cavity center. As discussed in Section~B, WGM cavities do not emit vertically, and direct experimental determination of the radial field maximum is generally difficult. However, in near-field measurements this becomes possible owing to the evanescent coupling between the tip and the resonator. Figure~S8 (see Section IV of the Supplementary Information) shows the FFT spectra extracted from circumferences at $1426\,\mathrm{cm}^{-1}$, where two distinct peaks corresponding to the $(9,2)$ and $(14,1)$ modes are observed. The spatial evolution of their FFT amplitudes confirms that the $(9,2)$ mode is localized closer to the cavity center, consistent with its assignment as the second radial-order mode.

We also investigated the influence of the auxiliary cavity geometry on WGM excitation. While the resonance frequencies and mode numbers are determined solely by the disk resonator, the auxiliary cavity geometry significantly affects the coupling efficiency to individual WGMs through the spatial distribution of the launched hyperbolic phonon-polaritons. Additional numerical simulations supporting this conclusion are presented in the Section IA of the Supplementary Information (Fig.~S2).

Similar experimental data are presented in the Supplementary Information (see Fig.~S9) for a 0.6-$\mu$m-radius, 26-nm-thick hBN disk with an auxiliary triangular cavity. The same behavior is observed; however, owing to the lower edge smoothness of this cavity, the corresponding WGM quality factors are reduced.

\vspace{-1.5em}
\section{Numerical Simulations}
\vspace{-1em}

3D finite element analysis allowed us to model the full variety of modes excited in the disk. These simulations confirm the principle of decoupled excitation via the auxiliary cavity and demonstrate good agreement between the calculated spatial field distributions and the s-SNOM measurements. To study each mode separately, we analyze the eigenmodes of the hBN microdisk and then compare the spectral and spatial features of selected modes with the experimental data.

We employ the effective index approximation (EIA) to evaluate the eigenmodes of the hBN disk. The EIA has been widely used to study various HPhP modes in structured hBN~\cite{Tamagnone2018,Orsini2024,borodin2026cavity} and simplifies the 3D problem by reducing it to a 1D problem for $z$-confinement in the layered structure and a 2D problem for the laterally structured hBN. While the experimental samples contain a $\sim 50$ nm Au layer between the hBN layer and the $\mathrm{SiO_2}$ substrate, the permttivity of Au is sufficiently negative in the frequency range of interest so that we may approximate it as a perfect electric conductor (PEC). First, we solve the 1D problem to obtain the HPhP modes of an air/hBN/PEC layered structure and determine the effective index, $n_{\mathrm{eff}}$. Solving the wave equation for TM modes in the layered structures with appropriate boundary conditions yields a transcendental dispersion equation for the HPhP modes
\begin{equation}
\begin{aligned}
k_0 d \sqrt{\epsilon_x\left(1-\frac{n_{\mathrm{eff},l}^2}{\epsilon_z}\right)}
&= \arctan\!\left(
\sqrt{\frac{\epsilon_x \epsilon_z (n_{\mathrm{eff},l}^2 - 1)}
{\epsilon_z - n_{\mathrm{eff},l}^2}}
\right) \\
&\quad + l\pi .
\end{aligned}
\label{eq:4}
\end{equation}

where $l$ is the order of the HPhP modes in the layered structure and $n_{\mathrm{eff},l}$ is the corresponding effective refraction index. Eq. \ref{eq:4} demonstrates the existence of multiple HPhP modes as well as the explicit dependence of $n_{\mathrm{eff},l}$ on the dispersive permittivity of hBN ($\epsilon_x,\ \epsilon_z$) and the thickness ($d$) of the hBN layer. The dispersion relations for the HPhP modes calculated from Eq.~\ref{eq:4} for different values of $l$ in the frequency range of interest (RSII) are shown in Fig.~S10b. Even for the lowest-order mode ($l=0$), we find $n_{\mathrm{eff},0} \sim 20$, highlighting the deeply subwavelength nature of the HPhPs. The modes exhibit finite loss, with $\mathrm{Im}(n_{\mathrm{eff}})/\mathrm{Re}(n_{\mathrm{eff}}) < 0.15$. Since the decay constant is much smaller than the propagation constant, we neglect $\mathrm{Im}(n_{\mathrm{eff}})$ in the subsequent calculations.

Next, we solve the 2D problem of a disk with refractive index $n_{\mathrm{eff}}$ embedded in a dielectric (air) environment to obtain the HPhP modes of the structured hBN.
\onecolumngrid
\noindent
\begin{center}
\includegraphics[width=1\textwidth]{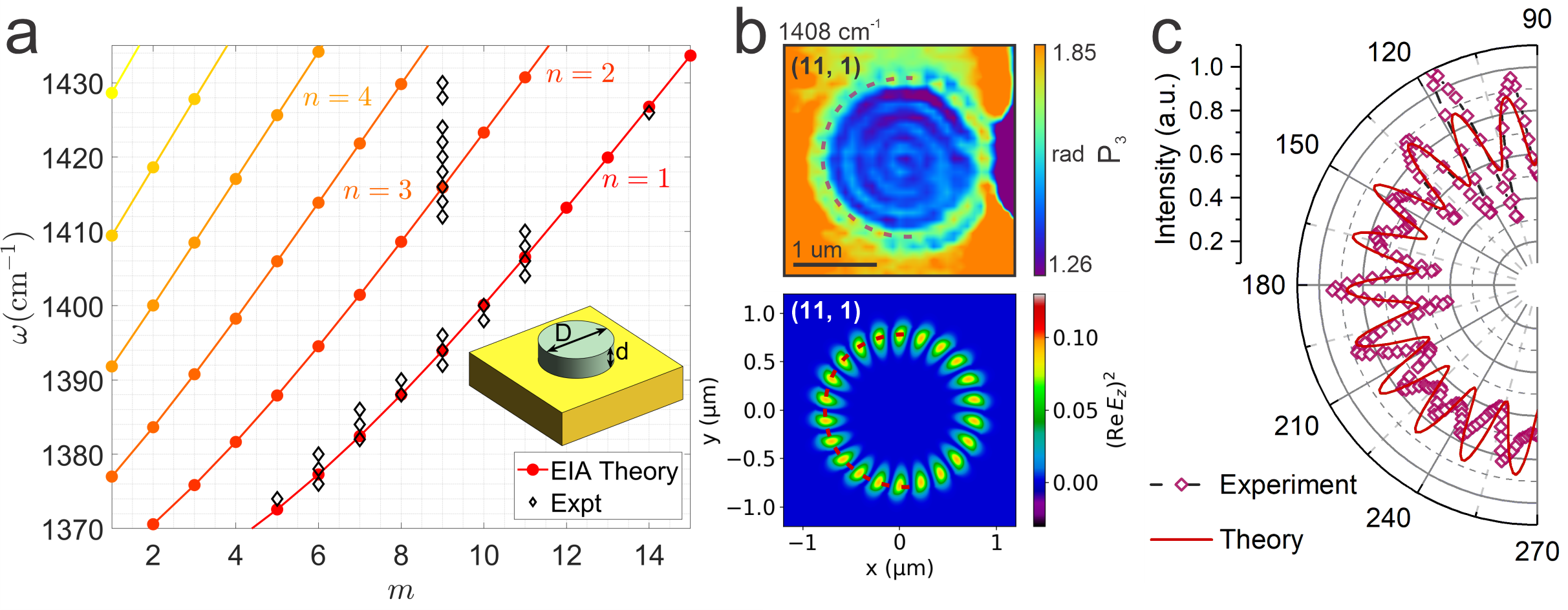} \\[1ex]
\parbox{1\textwidth}{\setlength{\parindent}{0pt}%
\textbf{Figure 4: Numerical eigenmodes of an hBN disk and comparison with experiment.}  
\textbf{a}) Azimuthal mode number $m$ as a function of wavenumber. Colored circles and solid lines denote mode positions calculated using the effective index approximation (EIA), and black open diamonds indicate experimental data.
\textbf{b}) Experimental s-SNOM image acquired at 1408~cm$^{-1}$ (top) and the corresponding field distribution, $(\mathrm{Re}\,E_z)^2$, obtained from a 2D EIA-based eigenmode solver (bottom).  
\textbf{c}) Experimental azimuthal profile extracted along the raspberry dashed line in \textbf{b} (raspberry diamonds) compared with the simulated profile extracted along the red dashed line in \textbf{b} (solid red line).

}
\end{center}
\twocolumngrid
\noindent
We use the standard solutions of the the wave equation in cylindrical coordinates and choose the $E_z\neq 0$ solution since that best matches the TM HPhP modes considered in the 1D solution. Enforcing the continuity of the tangential $E_z,\ H_{\phi}$ components at the disk boundary, we obtain,
\vspace{-1em}
\begin{equation}
    k_d \frac{J_{m+1}(k_d R)}{J_m(k_d R)} =
    k_0 \frac{K_{m+1}(k_0 R)}{K_m(k_0 R)} ,
    \label{eq:7}
\end{equation}
\vspace{0em}
where $J_m$ and $K_m$ are Bessel functions of the first kind and modified Bessel functions of the second kind, respectively, and $k_d=k_0n_{\mathrm{eff},0}$. Equation~\ref{eq:7} is a transcendental equation that must be solved together with Eq.~\ref{eq:4} to obtain the $\omega$--$m$ dispersion relation of the HPhP WGM modes in the hBN disk. For a given value of $m$, multiple solutions of Eq.~\ref{eq:7} can exist at different frequencies, corresponding to eigenmodes with different numbers of radial anti-nodes ($n$). A detailed description of the EIA formalism is provided in Section~VI, \textit{Effective Index Approximation for HPhP WGMs}, of the Supplementary Information.

Figure~4a shows the results of the above-discussed approach. The model captures not only different azimuthal mode numbers but also several radial mode orders within the investigated range. Hollow black diamonds represent the data extracted from the experiment. We note that the experimental results are in good agreement with the theory in terms of both azimuthal and radial mode numbers. Moreover, the theory predicts the coexistence of several low-$m$ modes together with the dominant high-$m$ mode. This behavior is also observed experimentally in the form of several weaker peaks in the FFT spectra (see Fig.~3d). In addition, at higher wavenumbers (e.g., around 1428~cm$^{-1}$), the experimental data approach the $n=3$ branch. Although a clear switching of the dominant mode is not observed experimentally, which we attribute to the high losses of modes with $n=3$, a weak peak near $(m,n)\approx(8,3)$ can still be detected at 1430~cm$^{-1}$. This observation is also in good agreement with the predictions of the EIA theory. 

It is important to note that the effective-index eigenmode solver predicts only the discrete resonance frequencies and field distributions of the cavity eigenmodes. In contrast to the experiment, it does not describe the frequency-dependent excitation efficiency or the finite linewidth of each resonance, and therefore does not produce a continuous \((\omega,k_{\phi})\) spectrum. Instead, each calculated point corresponds to the resonant frequency of a mode with a particular pair of quantum numbers \((m,n)\). Because the cavity radius is fixed, the experimentally measured azimuthal wavevector is directly related to the mode number through \(k_{\phi}=m/R\), enabling a direct comparison between the calculated \(\omega(m)\) dependence and the experimentally measured \(\omega(k_{\phi})\). The experimentally observed plateaus and overlapping resonances arise from the finite linewidths of the cavity modes, whereas the eigenmode calculations represent the ideal lossless resonance positions.

Figure~4b compares the experimental s-SNOM image (top) with the 2D EIA-simulated field, $(\mathrm{Re}\,E_z)^2$, distribution of the $(11,1)$ mode (bottom), which is identified in both theory and experiment at approximately 1408~cm$^{-1}$. To enable a direct comparison between the experimental and simulated data, azimuthal profiles were extracted along the rim of the disk in the left half of the structure (90--270$^\circ$), where the experimental signal is less perturbed by the launching cavity. The extracted profiles are shown in angular coordinates in Fig.~4c. As can be seen, the normalized experimental intensity profile (raspberry-colored hollow diamonds) is in good agreement with the EIA-simulated profile (red solid line) of the $(11,1)$ mode, further confirming the validity of the employed models and the interpretation of the experimental data.

\vspace{-1em}
\section{Discussion}
\vspace{-1em}

\subsection{On the scattering bottleneck in etching-defined cavities}
\vspace{-1em}
Despite extensive studies of polaritonic cavities, the dominant scattering bottleneck limiting the cavity quality factor remains unclear. In the polaritonics community, total cavity losses are commonly expressed as
$\
1/Q_{\mathrm{tot}} = 1/Q_{\mathrm{rad}} + 1/Q_{\mathrm{mat}} + 1/Q_{\mathrm{scat}},
$\
corresponding to radiative, material, and scattering losses, respectively. Strategies to enhance $Q_{\mathrm{tot}}$ therefore focus either on cavity design to suppress radiative and scattering losses or on material optimization to reduce intrinsic dissipation.

Etching-defined cavities suppress radiation leakage but introduce edge-related scattering, whereas dielectric-contrast-defined cavities reduce edge scattering at the cost of increased leakage\cite{kim2018photonic,herzig2024high,jackering2025tailoring,borodin2026cavity,chaudhary2019polariton}. Both approaches typically yield quality factors of a few hundred, while substantially higher values remain challenging. Even in this work, where we demonstrate WGMs that traditionally exhibit record-high Q factors among resonators, the achieved Q factor is still only comparable to the best values reported so far for HPhPs and does not exceed them. Although isotopically purified hBN is known to exhibit reduced material losses,\cite{giles2018ultralow} the dominant loss channel in realistic cavity geometries remains unresolved. To specifically investigate scattering mechanisms, we additionally fabricated several cavities from isotopically pure h$^{11}$B$^{14}$N. Given the inherent complexity of data analysis for WGM modes, we focused on simpler round-trip modes in hBN disks on SiO$_2$/Si substrates. The corresponding data is shown in Supplementary Information Section~VIII. Our findings show that edge scattering is the dominant scattering mechanism in etching-defined cavities, leading to temperature-independent Q factors for sufficiently thick cavities.

\vspace{-1em}
\subsection{Perspectives}
\vspace{-1em}

The direct observation and controlled excitation of hyperbolic phonon-polariton WGMs establish a general platform for exploring anisotropic polaritonic physics beyond the specific material system studied here. The excitation mechanism is broadly applicable to other in-plane anisotropic materials, such as MoO$_3$\cite{hu2020phonon}, and to twisted anisotropic heterostructures\cite{zhang2021hybridized}, where twist angle provides an additional handle to engineer in-plane anisotropy, dispersion, and mode structure.

Hyperbolic polariton WGMs offer a particularly attractive route to non-Hermitian physics in deeply subwavelength resonators. In-plane anisotropy breaks rotational symmetry and is expected to lift the degeneracy between clockwise and counterclockwise WGMs. Together with intrinsic losses and tunable mode coupling in anisotropic or twisted systems (such as twisted MoO$_3$)\cite{chen2020configurable}, this enables access to non-Hermitian regimes, including exceptional points, where eigenmodes and eigenvalues coalesce \cite{li2023exceptional}.

More broadly, anisotropy-induced deformation of hyperbolic WGM profiles enables directional energy flow and enhanced field confinement, opening opportunities for nanoscale implementations of transformation-optics concepts\cite{kim2016designing}. Combining extreme subwavelength confinement, dispersion engineering, and intrinsic dissipation, hyperbolic polariton WGMs provide a versatile testbed for studying non-Hermitian wave physics, dispersion topology, and ultrastrong light–matter interactions, with potential impact on on-chip nanophotonics, infrared spectroscopy, and polaritonic device technologies.

\vspace{-1.5em}
\section{Conclusion}
\vspace{-1em}

In conclusion, we introduce an auxiliary-cavity-assisted near-field excitation strategy that overcomes a fundamental limitation of conventional s-SNOM. While the scanning tip continues to both excite and detect phonon-polaritons, a resonant auxiliary cavity formed at the Au/SiO$_2$ interface provides an additional tip-independent excitation channel. Owing to its resonant nature, the auxiliary cavity launches hyperbolic phonon-polaritons with sufficiently high efficiency to produce pronounced excitation of whispering-gallery modes, while the scanning tip primarily probes the resulting near-field distribution.

Using this approach, we directly visualize hyperbolic whispering-gallery modes in deeply subwavelength hBN resonators with high quality factors and large, discrete azimuthal momentum ($k_\phi/k_0$ up to 15). Spatially resolved near-field imaging combined with Fourier analysis reveals the frequency-dependent excitation spectra and evolution of the excited whispering-gallery modes, enabling direct determination of their azimuthal and radial mode numbers. By tuning the excitation frequency, we demonstrate controlled switching between WGMs with different azimuthal and radial indices. Within the linewidth of a given mode, the azimuthal mode number remains unchanged while the effective refractive index evolves continuously before the resonator switches to the neighboring whispering-gallery mode.

Our results establish a minimally perturbative approach for probing the intrinsic field distributions of hyperbolic whispering-gallery modes. More broadly, the demonstrated excitation strategy provides a versatile platform for studying and engineering confined hyperbolic polaritons and opens new opportunities for manipulating high-angular-momentum polaritonic states and developing hyperbolic nanophotonic devices.

\section*{\textsf{Methods}}
\small
\subsection*{Device fabrication.}
\noindent
Si$^{++}$/SiO$_2$ substrates were pre-patterned using standard photolithography and metal deposition procedures. For photolithography, SUSS MicroTec MJB4 mask aligner and SPR-700 photoresist were used. After the photoresist development, we deposited metal (Ti/Au, 10/40 nm) using the Angstrom electron beam vacuum deposition system.

hBN flakes were prepared via mechanical exfoliation using PDMS adhesive tapes and were transferred on Si/SiO$_2$ substrates for selection. Relying on optical contrast, we selected flakes of suitable thicknesses and transferred them on the pre-patterned Si$^{++}$/SiO$_2$ substrates using the PMDS/PC dry transfer method\cite{wang2013one,purdie2018cleaning}.

For the cavity fabrication, we used the Raith EBPG5200 electron beam lithographer and PMMA 950 A4 e-beam resist. After the resist development, exposed hBN was etched out using Oxford ICP-RIE plasma etcher with $SF_6+O_2$ plasma source (40/10 sscm, 90 mTorr, 60 W).
\subsection*{Near-field optical measurements.}
\noindent
We used a commercially available scattering-type scanning near-field microscope (s-SNOM) developed by Neaspec/Attocube (cryo-neaSCOPE) with a pseudo-heterodyne mode to a detect a background-free optical response ($\Omega \pm NM$) using the decoupled optical amplitude and phase\cite{ocelic2006pseudoheterodyne, moreno2017phase,keilmann2004near}. PtIr-coated AFM tips (Nanoworld, 23 nm coating) with a resonant frequency $\approx$ 250 kHz were used to obtain the topography and the optical maps. AFM tips were illuminated by a tunable commercially available mid-infrared quantum cascade laser (QCL) using 1-2 mW of power (MIRcat, $\lambda$ = 5.5-11.5 $\mu$m from the DRS Daylight Solutions Inc.) All optical measurements presented in the manuscript were obtained on the 3rd or the 4th harmonics (i.e., S$_3$ and S$_4$, respectively) to ensure the background-free detection.

\bibliographystyle{naturemag-sergi}
\bibliography{hBN.bib}

\vspace*{-0.2cm}
\section*{\textsf{Acknowledgements}}
\vspace*{-0.2cm}
\noindent
P.S. acknowledges support from the start-up fund provided by the University of Notre Dame (\#373837). K.W. and T.T. acknowledge support from the JSPS KAKENHI (Grant Numbers 21H05233 and 23H02052) and World Premier International Research Center Initiative (WPI), MEXT, Japan. The work at Cornell was supported by the University of Dayton Research Institute (UDRI) under the contract FA8651-24-F-B013, Office of Naval Research (ONR) under the grant no. N00014-21-1-2056, and the Army Research Office (ARO) under the award W911NF2110180. S.L acknowledge support National Research Foundation of Korea (NRF) grant funded by the Korean government (MSIT) (NRF-RS-2023-NR076916). JHE and TP thank the National Science Foundation for award number 2413808, which supported the boron-enriched hBN crystal growth.

\vspace*{-0.2cm}
\section*{\textsf{Author contributions}}
\vspace*{-0.2cm}
{\small
\noindent
P.S., B.B., and H.H.S. conceived and designed the experiment. B.B. fabricated the samples and performed the measurements. Data were analyzed and interpreted by B.B., H.H.S., and P.S. S.M. worked on the effective index theory, S.R. and S.L. worked on the 3D driven simulations under the supervision of G.S. The theory and simulation work was done in consultation with B.B., H.H.S. and P.S. Naturally abundant hBN crystals were provided by K.W. and T.T., and isotopically pure h$^{11}$B$^{14}$N crystals were provided by J.H.E. and T.P. B.B. wrote the manuscript, and all co-authors contributed to editing. P.S. supervised the work.
}

\vspace*{-0.2cm}
\section*{\textsf{Competing Financial Interests}}
\vspace*{-0.2cm}
{\small
\noindent
The authors declare no competing financial interests.
}

\vspace*{-0.2cm}
\section*{\textsf{Data Availability Statement}}
\vspace*{-0.2cm}
{\small
\noindent
The data that support the findings of this study is available from the corresponding authors upon reasonable request.
}
\newpage
\end{document}


\title{\Large\textsf{\papertitle}}

\author{Bogdan Borodin}
\email{bborodin@nd.edu}
\affiliation{\footnotesize Department of Physics and Astronomy, University of Notre Dame, Notre Dame, IN 46556,~USA}
\affiliation{\footnotesize Stavropoulos Center for Complex Quantum Matter, University of Notre Dame, Notre Dame, IN 46556,~USA}

\author{Samyobrata Mukherjee}
\affiliation{\footnotesize School of Applied and Engineering Physics, Cornell University, Ithaca, NY 14853, USA}

\author{Shivaksh Rawat}
\affiliation{\footnotesize School of Applied and Engineering Physics, Cornell University, Ithaca, NY 14853, USA}

\author{Seojoo Lee}
\affiliation{\footnotesize School of Applied and Engineering Physics, Cornell University, Ithaca, NY 14853, USA}
\affiliation{\footnotesize The Institute of Basic Science, Korea University, Seoul 02841, Republic of Korea}

\author{Thomas Poirier}
\affiliation{\footnotesize  Tim Taylor Department of Chemical Engineering, Kansas State University, Manhattan, 66506,
Kansas, ~USA}

\author{Kenji Watanabe}
\affiliation{\footnotesize Research Center for Electronic and Optical Materials, National Institute for Materials Science, 1-1 Namiki, Tsukuba 305-0044,~Japan}

\author{Takashi Taniguchi}
\affiliation{\footnotesize Research Center for Materials Nanoarchitectonics, National Institute for Materials Science,  1-1 Namiki, Tsukuba 305-0044,~Japan}

\author{James H. Edgar}
\affiliation{\footnotesize  Tim Taylor Department of Chemical Engineering, Kansas State University, Manhattan, 66506,
Kansas, ~USA}

\author{Hanan Herzig Sheinfux}
\affiliation{\footnotesize  Department of Physics, Bar-Ilan University, Ramat Gan 52900, Israel}

\author{Gennady Shvets}
\affiliation{\footnotesize School of Applied and Engineering Physics, Cornell University, Ithaca, NY 14853, USA}

\author{Petr Stepanov}
\email{pstepano@nd.edu}
\affiliation{\footnotesize Department of Physics and Astronomy, University of Notre Dame, Notre Dame, IN 46556,~USA}
\affiliation{\footnotesize Stavropoulos Center for Complex Quantum Matter, University of Notre Dame, Notre Dame, IN 46556,~USA}

\keywords{phonon-polariton, polaritonic optics, strong confinement, hexagonal boron nitride, focusing}


\begin{abstract}
\vspace*{0.2cm}

\end{abstract}

\maketitle

\renewcommand{\figurename}{Fig.}
\setcounter{equation}{0}
\setcounter{figure}{0}
\setcounter{table}{0}
\makeatletter
\renewcommand{\theequation}{S\arabic{equation}}
\renewcommand{\thefigure}{S\arabic{figure}}
\renewcommand{\thetable}{\arabic{table}}
\renewcommand{\bibnumfmt}[1]{[#1]}
\renewcommand{\citenumfont}[1]{#1}

\tableofcontents
\onecolumngrid
\newpage
\section{3D driven simulations of the \( |E_z|^2 \) field distribution in the disk--cavity structure}
To numerically model the excitation of whispering-gallery modes in hBN nanodisks, we performed 3D frequency-driven simulations using a commercially available electromagnetics solver, COMSOL Multiphysics. We defined a computational domain of size $5\ \mu\mathrm{m\ (x)}\times3\ \mu\mathrm{m\ (y)}\times 5\ \mu\mathrm{m\ (z)}$, where the hBN nanodisk and the hBN cavity are placed in the x-y plane and on top of the silica substrate of thickness 2 $\mu \mathrm{m}$. The circular hBN nanodisk of 32 nm thickness is placed on a gold film of 50 nm thickness, which is then placed on the silica substrate. However, the hBN cavity that launches the whispering-gallery modes into the circular nanodisk is placed on top of the bare silica substrate. This creates a difference in the z-locations of the cavity and the nanodisk and results in the bending of the composite nanodisk and cavity structure. Therefore, to emulate the bending of the hBN film, we connect the hBN nanodisk on the gold film with the hBN cavity on the bare silica substrate with a ramp structure made of hBN only. The diameter of the simulated hBN nanodisk is taken to be slightly smaller ($\sim 1.7\ \mu m$) than the fabricated one ($\sim 1.9\ \mu m$) due to the smaller effective diameter of the nanodisk resulting from the rough edges of the fabricated nanodisk. The material properties of hBN and silica ($\mathrm{SiO_2}$) are taken from refs.~\cite{kumar2015tunable,kischkat2012mid}. The gold film is modeled using a Drude-Lorentz dispersion model with a plasma frequency of 9 eV, and the electronic losses are modeled using a scattering frequency of 0.17 eV. To simulate plane-wave incidence onto the hBN cavity, we modeled the wall in the y-z plane close to the hBN cavity in the computational domain as a constant current-density surface that excites an electromagnetic wave linearly polarized in the z direction. PMLs are placed on the top and bottom of the simulation domain for better convergence of the numerical simulations. The results obtained are shown in Fig.~3b.

\begin{figure}[H]
    \centering
    \includegraphics[scale=0.9]{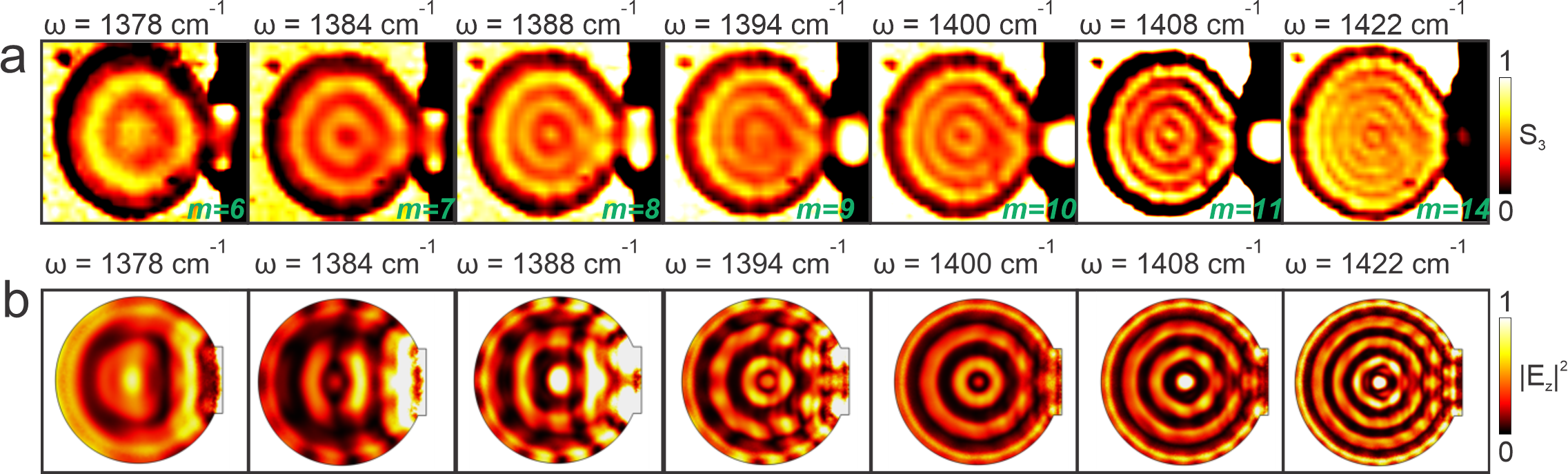}
   \caption{
    \textbf{\( |E_z|^2 \) vs S$_3$} a) Experimental s-SNOM amplitude demodulated at the third harmonic (S$_3$); b) 3D driven simulations \( |E_z|^2 \) field distribution in the disk--cavity structure at various excitation wavenumbers correcponsing to shown in the main text. 
}
    \label{fig:SI3}  
\end{figure}

Figure~\ref{fig:SI3} demonstrates the comparison of the \( |E_z|^2 \) field distribution in the disk--cavity structure obtained via 3D driven simulations and s-SNOM amplitude (S$_3$) maps. 

\newpage
\subsection{Role of the auxiliary cavity geometry}

To investigate the influence of the auxiliary cavity geometry on WGM excitation, we performed three-dimensional driven simulations of the $|E_z|^2$ field distribution for four different launching configurations: (i) the rectangular cavity used throughout the main text, (ii) a triangular cavity, (iii) a localized defect at the Au interface that breaks the azimuthal symmetry, and (iv) a disk resonator without an auxiliary cavity or other launching structure. The simulated geometries and corresponding field distributions are shown in Fig.~S2.

\vspace{-1.5em}
\begin{figure}[h!]
    \centering
    \includegraphics[width=1.0\linewidth]{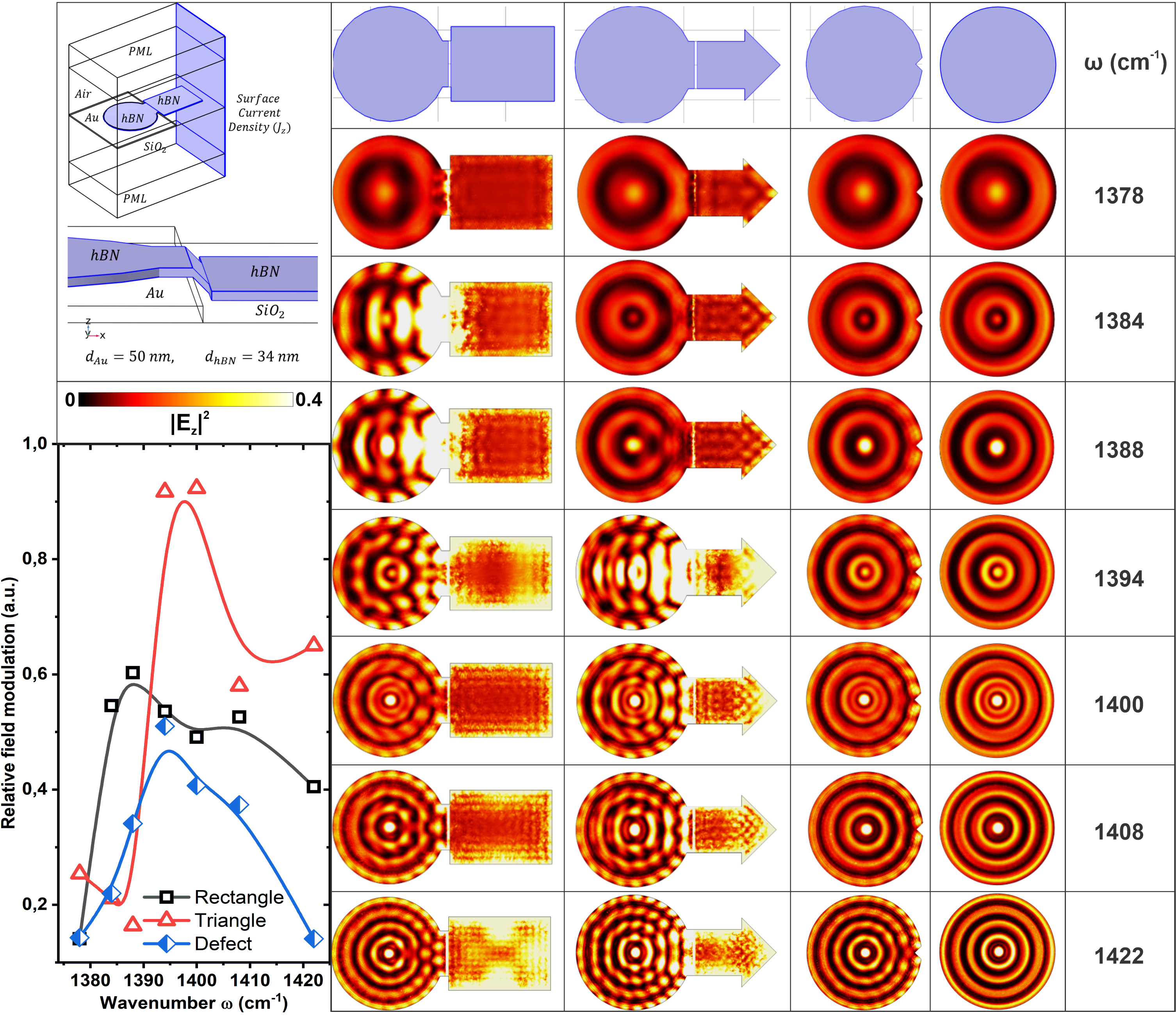}
    \vspace{-1.5em}
    \caption{Influence of the auxiliary cavity geometry on the excitation of whispering-gallery modes.}
    \vspace{-1.5em}
    \label{fig:SI_shape}
\end{figure}

As shown in Fig.~S2, the structure of the whispering gallery modes is not revealed in the absence of an auxiliary cavity, demonstrating that a resonant launching structure is essential for efficient WGM excitation under far-field illumination. In contrast, all three auxiliary-cavity geometries support WGM excitation, although with markedly different efficiencies. For example, the rectangular cavity provides stronger coupling at lower wavenumbers, whereas the triangular cavity becomes more efficient at higher wavenumbers. The localized defect also excites WGMs, but with substantially lower efficiency throughout the investigated spectral range.

These results support two important conclusions. First, the auxiliary cavity provides a resonantly enhanced launching mechanism that is essential for efficient excitation of whispering-gallery modes. Second, while the resonance frequencies and allowed mode numbers are determined solely by the disk resonator, the geometry of the auxiliary cavity modifies the angular distribution of the launched hyperbolic phonon-polaritons and therefore the overlap with the whispering-gallery modes, resulting in geometry-dependent coupling efficiencies.

\subsection{Role of the Au/SiO$_2$ interface}
\vspace{-1.0em}
Another essential component of the proposed excitation scheme is the Au/SiO$_2$ interface. It serves two distinct purposes. First, the impedance mismatch at the interface forms a partially reflecting boundary that spatially separates the auxiliary cavity from the disk resonator while still allowing efficient coupling between them. Second, the Au edge efficiently couples the incident far-field radiation into hyperbolic phonon-polaritons, providing the initial excitation of the auxiliary cavity. To verify these two roles independently, we performed two additional numerical studies.

First, we removed the Au/SiO$_2$ interface by placing both the auxiliary cavity and the disk directly on the SiO$_2$ substrate while keeping all other geometrical parameters unchanged. The resulting field distribution is shown in Fig.~S3.
\vspace{-1.0em}
\begin{figure}[h!]
    \centering
    \includegraphics[width=1.0\linewidth]{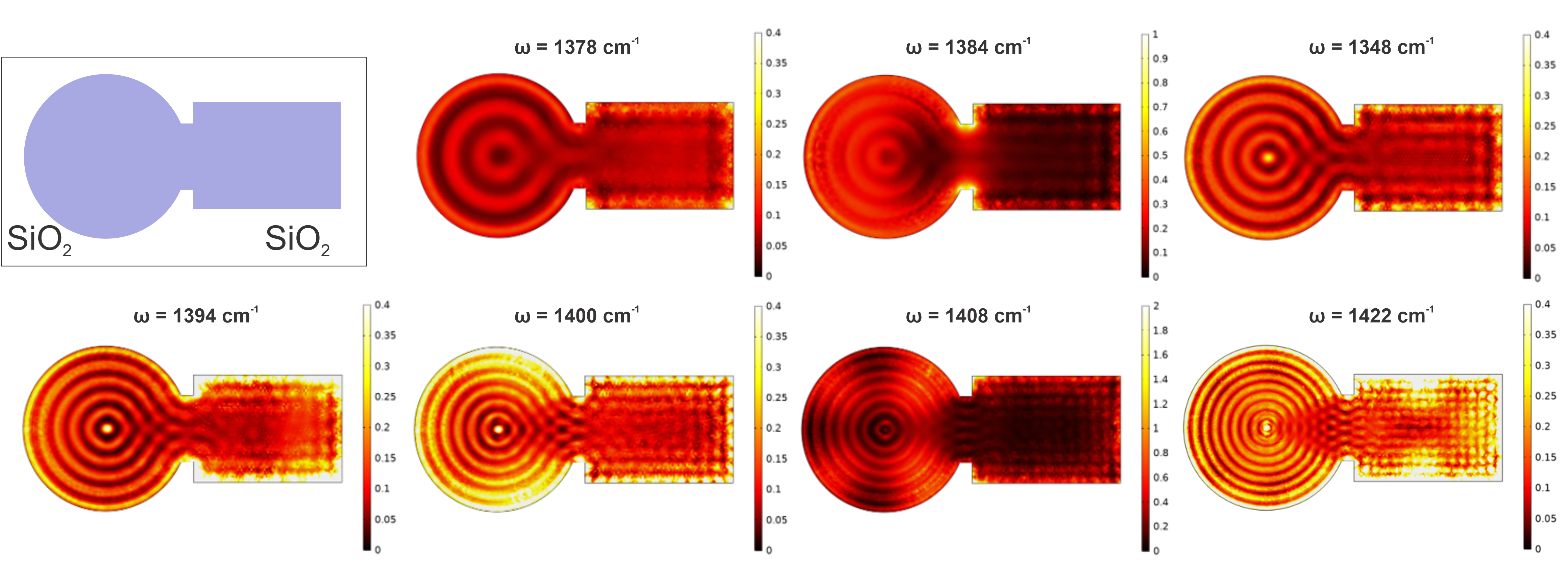}
    \vspace{-1.0em}
    \caption{Driven simulations after removing the Au/SiO$_2$ interface.}
    \label{fig:SI_interface}
\end{figure}
As shown in Fig.~S3, removing the impedance-mismatched interface causes the auxiliary cavity and the disk to become part of a continuous hBN sheet. Consequently, the auxiliary cavity no longer supports a localized resonant mode, while the disk is no longer isolated from the launching structure. Hyperbolic phonon-polaritons propagate freely between the two regions, preventing efficient confinement of both the launching cavity mode and the whispering-gallery modes. As a result, the characteristic WGM field distribution disappears almost completely. This demonstrates that the Au/SiO$_2$ interface is essential for forming two coupled but distinct resonators.

Second, we investigated the role of the Au edge in coupling the incident light into the auxiliary cavity. For this purpose, the Au/SiO$_2$ interface was restored, but the structure was illuminated from the substrate side. This geometry suppresses near-field scattering at the Au edge located in direct contact with the hBN layer. Figure~S4 schematically compares the top- and backside-illumination geometries and presents the corresponding simulated field distributions under backside illumination.
\begin{figure}[h!]
\centering
\includegraphics[width=1.0\linewidth]{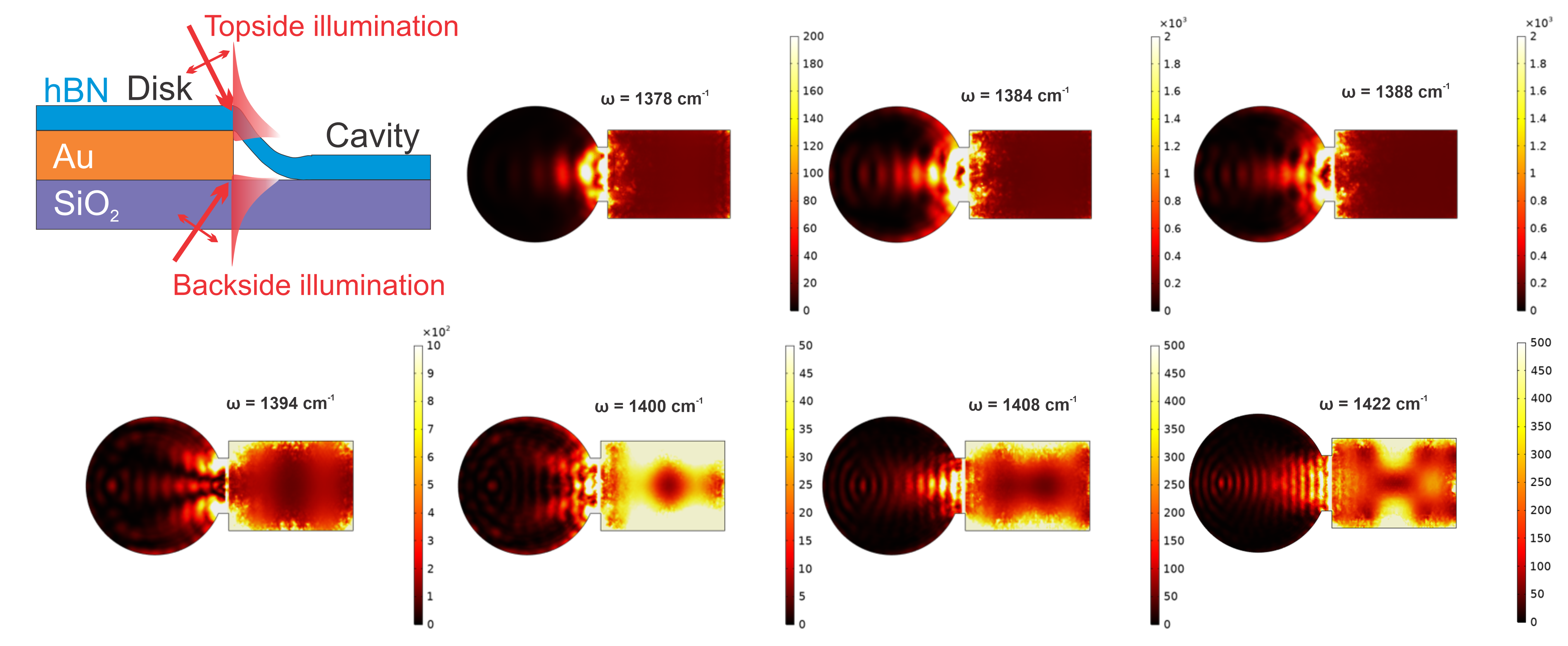}
\caption{Driven simulations of backside illumination in the presence of the Au/SiO$_2$ interface.}
\label{fig:backside}
\end{figure}
As shown in Fig.~S4, weak whispering-gallery modes can still be identified under backside illumination; however, their excitation efficiency is dramatically reduced compared with the top-illumination geometry. This behavior originates from the different scattering geometries. Under top illumination, the incident field is scattered at the Au edge located in direct contact with the hBN layer, efficiently coupling the far-field radiation into the localized resonance of the auxiliary cavity. In contrast, under backside illumination, scattering occurs at the buried Au/SiO$_2$ interface, which is spatially separated from the suspended hBN bridge connecting the auxiliary cavity to the disk. Consequently, only a weak scattered field reaches the auxiliary cavity, resulting in inefficient excitation of the launcher and therefore much weaker coupling to the whispering-gallery modes.

Together with the simulations presented in Fig.~S3, these results demonstrate that the Au/SiO$_2$ interface plays two indispensable roles in the proposed excitation scheme. The impedance-mismatched interface isolates the auxiliary cavity from the disk resonator, enabling the formation of two coupled but distinct resonant systems and thereby allowing high-Q WGMs to develop. At the same time, the Au edge provides efficient coupling of the incident far-field radiation into the auxiliary cavity. Removing the interface suppresses the WGMs almost completely, whereas preserving the interface but eliminating efficient edge scattering leaves the WGMs only weakly excited.

\section{Amplitude and phase maps used for the analysis of mode switching in Fig.~3 of the main text}

Figure~S5 presents the complete set of s-SNOM optical phase maps acquired at the fourth harmonic ($S_4$) with a spectral step of 2~cm$^{-1}$ and used to analyze the evolution of the whispering-gallery modes shown in Fig.~3 of the main text. The phase maps provide enhanced visual contrast of the mode patterns and are therefore useful for qualitative inspection of the modal evolution.

\begin{figure}[h!]
    \centering
    \includegraphics[width=0.7\linewidth]{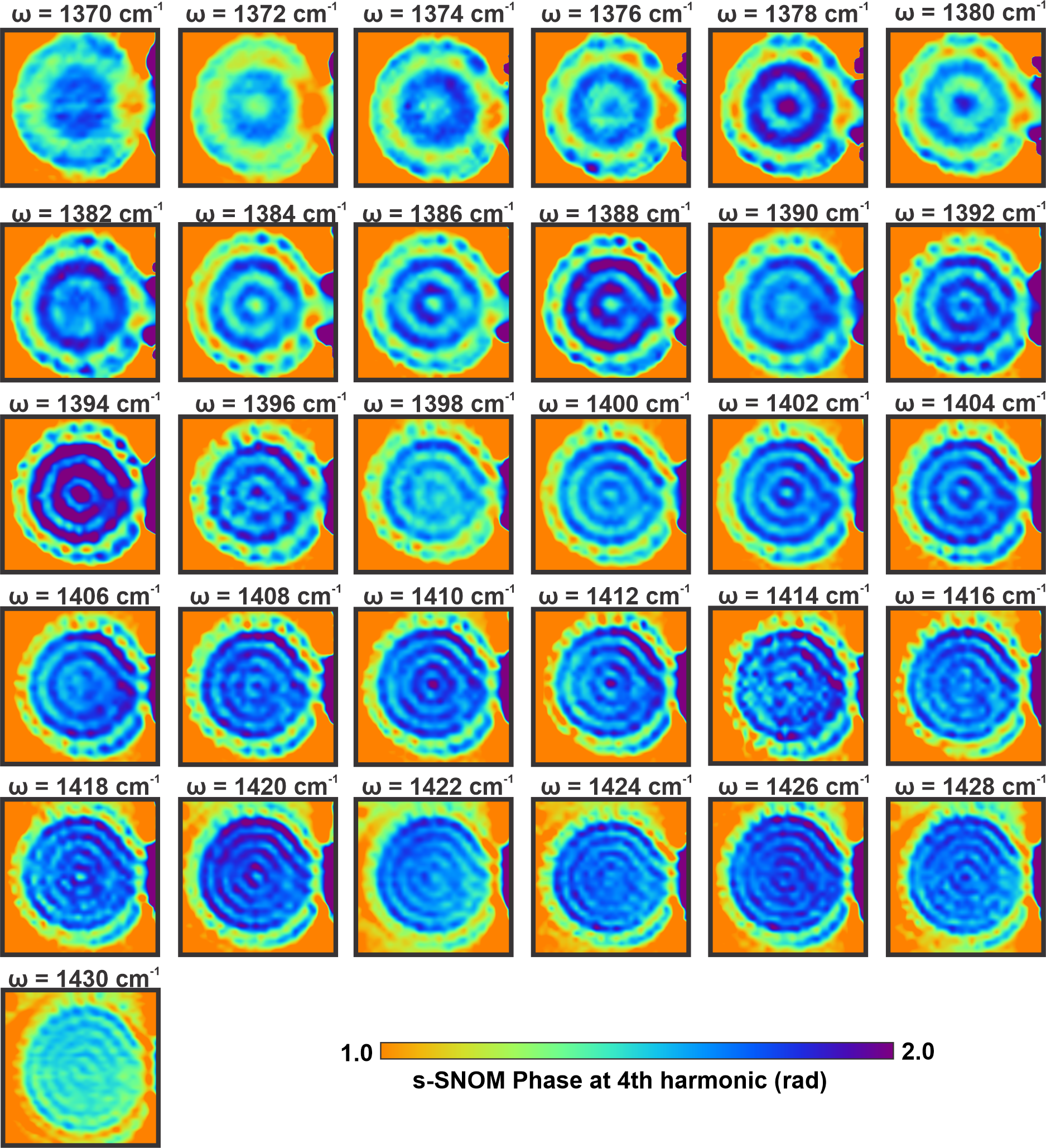}
    \caption{Complete set of s-SNOM optical phase maps ($S_4$) used to analyze the frequency evolution of whispering-gallery modes presented in Fig.~3 of the main text.}
    \label{fig:phase_maps}
\end{figure}

Figure~S6 presents the corresponding s-SNOM optical amplitude maps acquired under identical experimental conditions. As discussed in the main text, the FFT analysis was performed using the amplitude data because the phase uncertainty increases in regions of low near-field intensity, making the amplitude a more robust quantity for quantitative spectral analysis.

\begin{figure}[h!]
    \centering
    \includegraphics[width=0.7\linewidth]{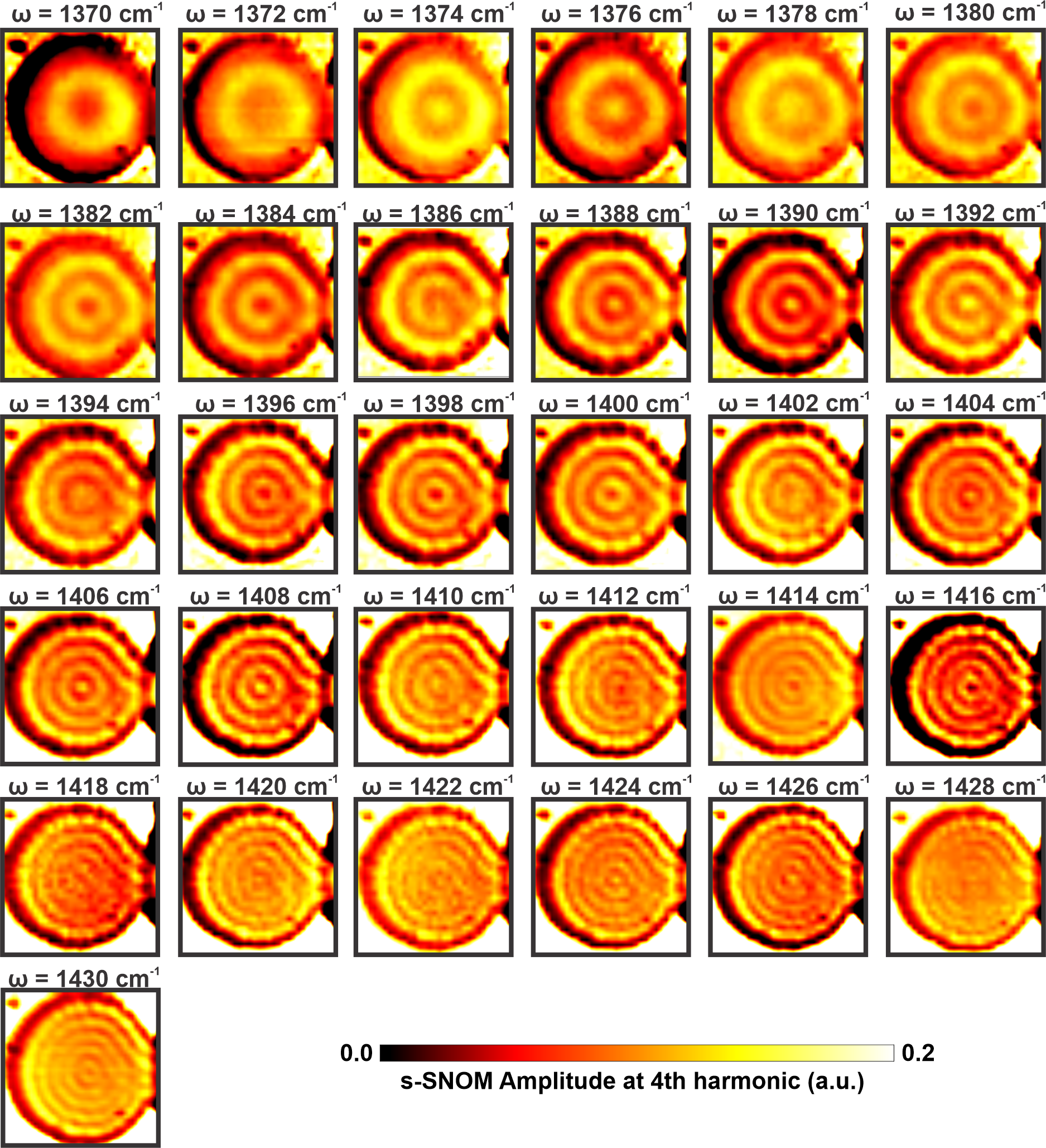}
    \caption{Complete set of s-SNOM optical amplitude maps ($S_4$) used for the FFT analysis presented in Fig.~3 of the main text.}
    \label{fig:amplitude_maps}
\end{figure}

It is important to emphasize that, although the optical phase maps in Fig.~S5 provide superior visual contrast of the whispering-gallery modes, all quantitative analysis in the main text was performed using the amplitude maps shown in Fig.~S6. In pseudo-heterodyne s-SNOM, the detected complex near-field signal is reconstructed from the in-phase and quadrature components, \(S=I+iQ\), with amplitude and phase given by \(A=\sqrt{I^{2}+Q^{2}}\) and \(\phi=\tan^{-1}(Q/I)\), respectively. The phase often appears visually cleaner because it is largely insensitive to slowly varying intensity backgrounds arising from changes in coupling efficiency, illumination intensity, and the round-trip contribution. As a result, the standing-wave pattern of the WGMs is more readily distinguished from the background.

However, the phase is intrinsically more susceptible to experimental noise. Propagation of the measurement uncertainty gives
\[
\delta A=\frac{I\,\delta I+Q\,\delta Q}{A},
\qquad
\delta\phi=\frac{I\,\delta Q-Q\,\delta I}{I^{2}+Q^{2}}
          =\frac{I\,\delta Q-Q\,\delta I}{A^{2}},
\]
demonstrating that the phase uncertainty scales approximately as \(1/A\). Consequently, in regions where the near-field signal is weak, even small fluctuations in the detected quadrature signals produce large phase variations. In contrast, the amplitude uncertainty remains approximately proportional to the detector noise and therefore provides a substantially more robust quantity for quantitative analysis. For this reason, although the phase maps are used throughout this work for visualization of the modal evolution, all Fourier-transform analysis and extraction of mode parameters were performed using the amplitude data.

\newpage
\section{Model-selection analysis for Lorentzian decomposition}
To evaluate whether the Lorentzian decomposition used to extract the WGM resonance parameters is statistically justified, we performed a model-selection analysis based on the Bayesian Information Criterion (BIC). The experimental pseudo-spectroscopic response was fitted with a variable number of Lorentzian components \(N\), ranging from \(N=3\) to \(N=25\). For each fit, the BIC was calculated as

\[
\mathrm{BIC}=k\ln(n)-2\ln(\hat{L}),
\]

where \(k\) is the number of free fitting parameters, \(n\) is the number of experimental data points, and \(\hat{L}\) is the maximum likelihood of the fitted model. Lower BIC values indicate a better balance between goodness of fit and model complexity.

\begin{figure}[h!]
    \centering
    \includegraphics[width=0.85\linewidth]{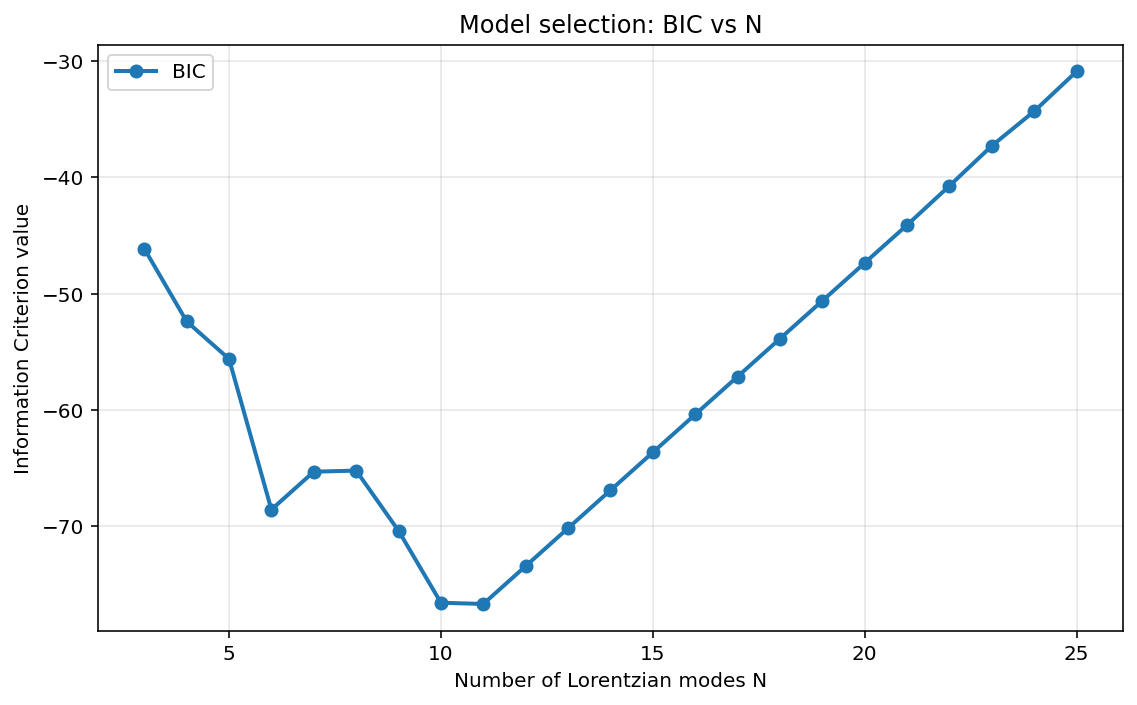}
    \caption{Bayesian Information Criterion (BIC) as a function of the number of Lorentzian components used to fit the pseudo-spectroscopic WGM response.}
    \label{fig:BIC}
\end{figure}

As shown in Fig.~\ref{fig:BIC}, the BIC decreases rapidly as the number of Lorentzian components increases up to \(N \approx 9\). The minimum is reached in the range \(N=10\)--\(11\), after which the BIC increases monotonically, indicating that additional Lorentzian components overfit the data. Therefore, the 9-Lorentzian model used in the main text lies within the statistically optimal region and captures the experimentally resolved resonance structure without introducing unnecessary degrees of freedom.

We also note that the 9-mode model is physically motivated by the number of experimentally visible resonances in the FFT-derived pseudo-spectrum and by the quasi-equidistant WGM mode sequence expected for the disk resonator. Fits with fewer than 9 Lorentzians leave structured residuals, indicating underfitting, whereas fits with more than 11 Lorentzians increase model complexity without improving the description of the data. The extracted resonance positions and linewidths remain stable within this optimal region, supporting the robustness of the reported Q-factors.

\newpage
\section{Observation of Mode Localization in the Multimodal Regime}
\begin{figure} [H]
    \centering
    \includegraphics[scale=0.9]{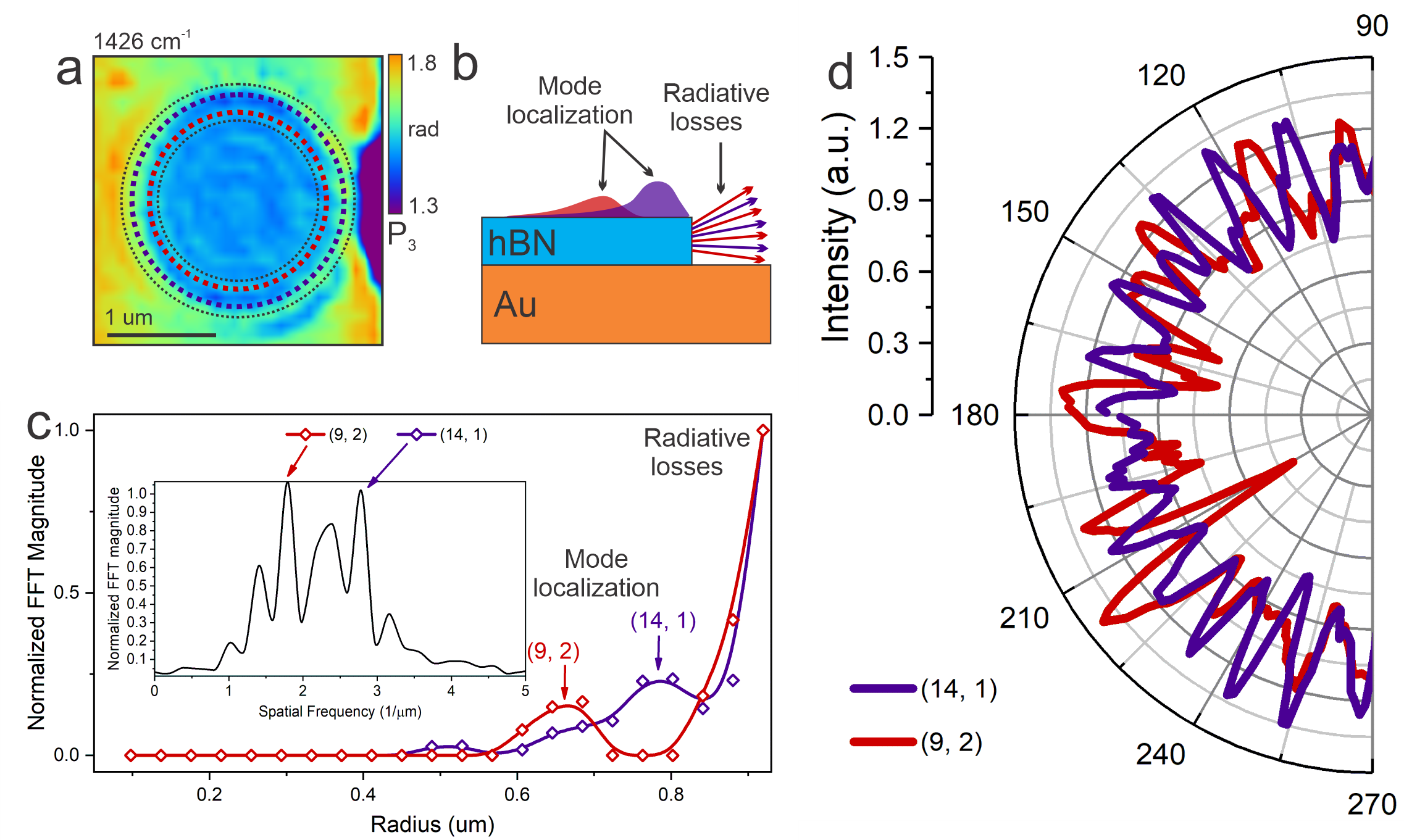}
    \caption{\textbf{Mode localization in the multimodal regime.}
    \textbf{a)} Near-field phase (P$_3$) map of a 1-$\mu$m-radius disk with an auxiliary cavity, recorded at $\omega = 1426~\mathrm{cm}^{-1}$.  
    \textbf{b)} Schematic illustration of mode localization in the multimode regime.  
    \textbf{c)} Analysis of the spatial-frequency distribution along the radius extracted from the selected arc profile.  
    \textbf{d)} Radial profiles taken at radii corresponding to the peak Fourier frequency inside the cavity.
    }
    \label{fig:SI1}  
\end{figure}

Using the approach described in the main text, spatial mode localization can be identified even in the multimode regime. Figure~\ref{fig:SI1} presents data acquired at $\omega = 1426~\mathrm{cm}^{-1}$, where a multimode regime is observed. Figure~\ref{fig:SI1}a shows a near-field phase (P$_3$) map of a 1-$\mu$m-radius disk with an auxiliary cavity. The black dashed lines indicate the range of radii over which radial profiles were summed and analyzed to obtain the FFT spectrum shown in the inset of Fig.~\ref{fig:SI1}c. The purple and red dashed lines mark the arcs used to extract profiles presented in Fig.~\ref{fig:SI1}d.

Figure~\ref{fig:SI1}b provides a schematic illustration of mode localization in the multimode regime. Two modes with different azimuthal ($m$) and radial ($n$) quantum numbers are localized near the cavity edge. The mode with higher $m$ and lower $n$ is pushed closer to the edge, whereas the mode with lower $m$ and higher $n$ is localized closer to the cavity center. At the same time, both modes experience radiative losses through the cavity edge (normal to the sidewall). Due to the coincidence of these radiative losses, both modes exhibit a pronounced peak in the FFT spectra at the cavity rim. However, the near-field technique allows access to the true spatial localization of the modes inside the cavity.

Figure~\ref{fig:SI1}c shows the radial distribution of FFT peaks corresponding to the (9, 2) and (14, 1) modes. Evidently, the (9, 2) mode is localized closer to the cavity center, whereas at the very rim both FFT components sharply increase for the reasons discussed above. The inset shows the FFT spectrum integrated over the radial range highlighted by the black dashed lines in Fig.~\ref{fig:SI1}a. After identifying the spatial-frequency peak of interest, we analyzed the magnitude of this selected component as a function of radius along the arc, from near zero up to distances exceeding the cavity radius.

Using the information on mode localization obtained from Fig.~\ref{fig:SI1}c, we extracted profiles along arcs at the corresponding radii to isolate the (9, 2) and (14, 1) modes. Complete separation of the modes is not possible due to their partial spatial overlap and the presence of interference effects. Nevertheless, Fig.~\ref{fig:SI1}d shows profiles predominantly corresponding to the (9, 2) and (14, 1) modes. While some noise and intersecting peaks are present, the red profile is in good agreement with the expected pattern of the (9, 2) mode, and the purple profile exhibits the number of peaks consistent with the (14, 1) mode.

These results demonstrate that the proposed approach enables not only identification of the spectral features of WGMs, but also investigation of their spatial characteristics, including determination of mode localization.

\section{Hyperbolic WGM$s$ in the 0.6-$\mu m$-radius, 26-$nm$-thick $h$BN disk with an auxiliary triangular cavity}

\begin{figure}[H]
    \centering
    \includegraphics[scale=0.9]{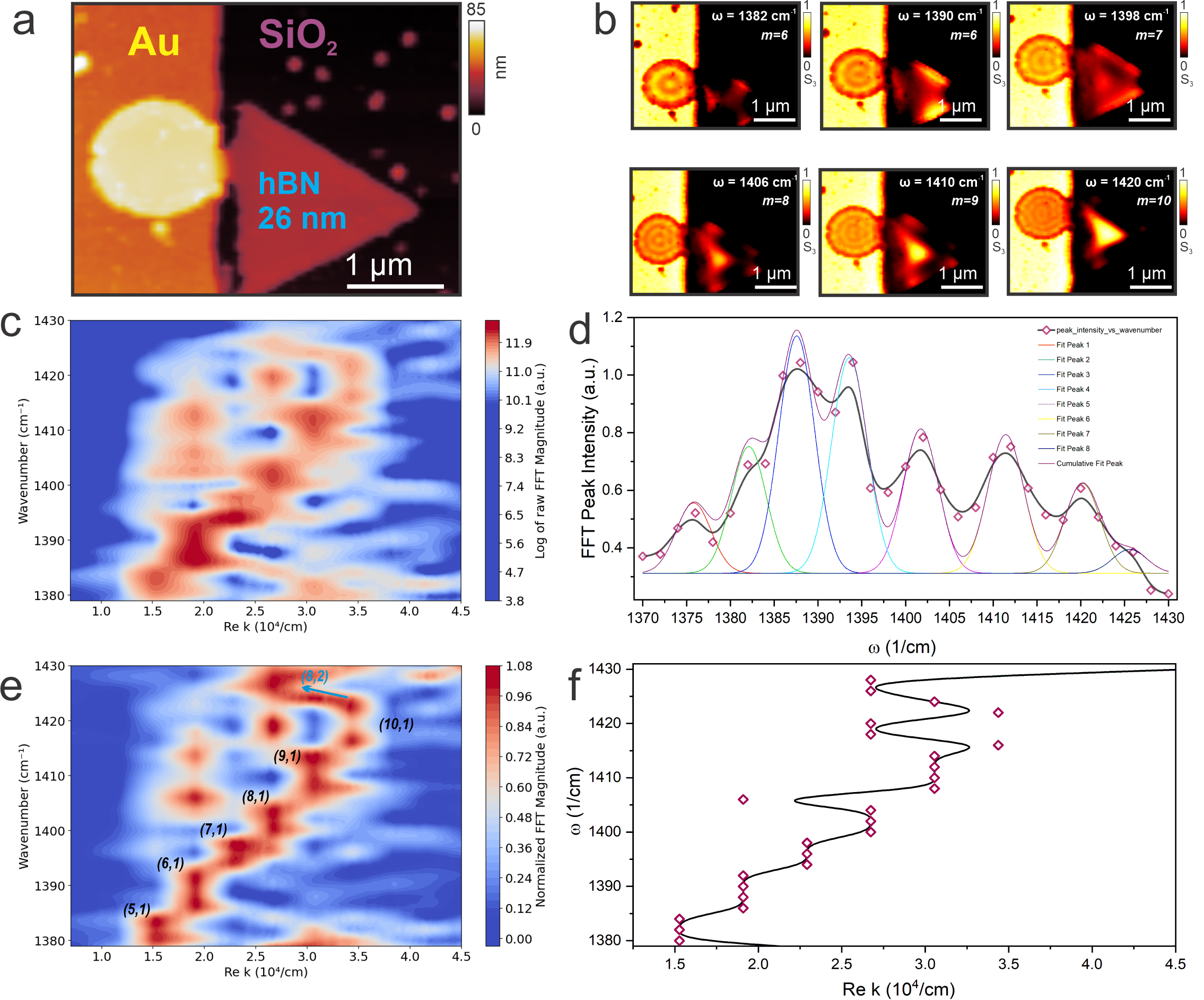}
   \caption{
    \textbf{0.6-$\mu$m-radius, 26-nm-thick hBN disk with an auxiliary triangular cavity.} 
    \textbf{a)} AFM topography of a 0.6-$\mu$m-radius, 26-nm-thick hBN disk positioned on top of a 50-nm-thick gold stripe, which is connected to a triangular hBN cavity resting on the SiO$_2$ substrate. 
    \textbf{b)} s-SNOM optical amplitude images at several $\omega$ between 1382–1420 cm$^{-1}$, demodulated at the third harmonic of the tip oscillation ($S_3$). 
    \textbf{c)} Contour plot constructed from raw FFT spectra of profiles along circumferences across excitation wavenumbers from 1370 to 1430 cm$^{-1}$, with a step size of 2 cm$^{-1}$. 
    \textbf{d)} Spectral positions of FFT maxima. The black curve represents a Lorentzian fit.  
    \textbf{e)} Contour plot constructed from normalized FFT spectra. Numbers in brackets indicate mode indices specific for peak position, $(m, n)$. 
    \textbf{f)} Dispersion curve. Raspberry-colored diamonds indicate the exact positions of FFT peaks. The mode numbers (extracted directly from maps) are shown in brackets.
}
    \label{fig:SI2}  
\end{figure}

We have studied several cavities of various diameters and auxiliary resonator shapes. In general, the physics described in the main text remains universal across these structures. Here, we present a 26-nm-thick, 0.6-µm-radius disk with an attached triangular auxiliary cavity. Figure~\ref{fig:SI2}a shows the AFM topography. Figure~\ref{fig:SI2}b presents s-SNOM amplitude maps at various $\omega$, demodulated at the third harmonic ($S_3$). As can be seen, the azimuthal modulation characteristic of WGMs is clearly visible on the maps. Using the approach described in the main text, the azimuthal mode numbers ($m$) were determined and indicated on the figures. Figure~\ref{fig:SI2}c shows a contour plot constructed from raw FFT spectra of profiles along circumferences. Figure~\ref{fig:SI2}d presents the spectral positions of maxima across the whole spectrum, with the black curve representing a Lorentzian fit. Figure~\ref{fig:SI2}e shows a contour plot constructed from normalized FFT spectra. Numbers in brackets indicate the mode indices corresponding to each peak, $(m, n)$. Figure~\ref{fig:SI2}f shows the dispersion curve. Raspberry-colored diamonds indicate the exact positions of FFT peaks, with the mode numbers (extracted directly from the maps) shown in brackets.

\vspace{-3.5em}
\section{Effective Index Approximation for HPhP WGMs}
We have studied the azimuthal structure of HPhP WGMs in a hBN microdisk by launching these HPhP WGMs using an auxiliary hBN cavity attached to the hBN disk that breaks azimuthal symmetry. While full wave 3D FEM simulations confirm the principle of decoupled excitation via the auxiliary cavity and show the correspondence of the calculated spatial field distributions to the s-SNOM measurements, we gain further insight into the nature of the modes that are excited by the auxiliary cavity by studying the eigenmodes of the hBN microdisk in the upper Reststrahlen band (RSII). We consider a hBN disk of thickness $d$ and diameter $D$ (radius $R=D/2$) placed on a perfect electric conductor (PEC) substrate. Though the experimental samples comprise a hBN disk on a gold layer of $\sim 50$ nm, the relative permittivity of gold for the frequencies of interest is extremely negative ($\epsilon_{\mathrm{Au}}(\omega\in \mathrm{RSII})<-1000$), and therefore we may approximate it with a PEC to simplify our calculations \cite{Ambrosio2018}.

We employ the effective index approximation (EIA) to calculate the eigenmodes of the hBN disk. EIA has been used to study a variety of HPhP modes in structured hBN \cite{Tamagnone2018, Orsini2024, borodin2026cavity} and it simplifies the 3D problem by splitting it into a 1D problem for layered structures and a 2D problem for the structured hBN. First, we solve the 1D problem to obtain the HPhP modes of an air/hBN/PEC layered structure to obtain the effective index ($n_{\mathrm{eff}}$). Then, we solve the 2D problem of a disk of refractive index $n_{\mathrm{eff}}$ embedded in a dielectric environment to obtain the HPhP modes of the structured hBN.
\vspace{-1.5em}
\subsection{HPhP modes in a layered air/hBN/PEC structure}
\label{sec:1D}
\begin{figure}[H]
    \centering
    \includegraphics[width=0.85\linewidth]{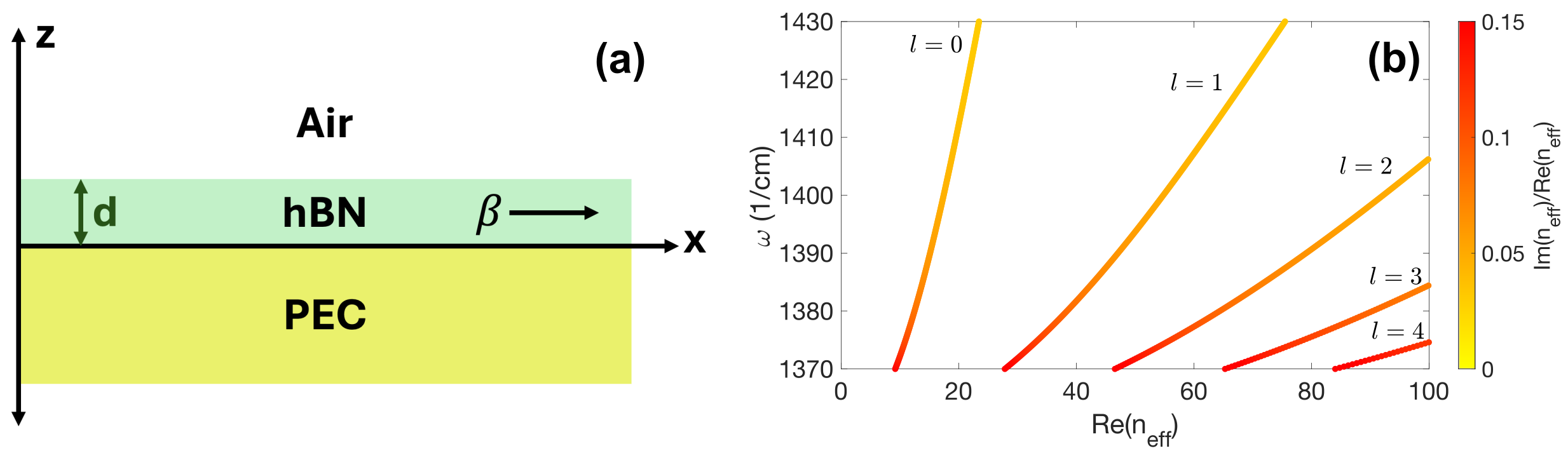}
    \caption{(a) Schematic of the layered air/hBN/PEC system with HPhPs propagating along $x$. (b) HPhP dispersion plots showing the effective index of the HPhPs for different values of $l$ when hBN thickness $d=30$ nm. The color shows the frequency dependent losses of the HPhP modes, specifically, the ratio $\mathrm{Im}(n_{\mathrm{eff}}) / \mathrm{Re}(n_{\mathrm{eff}})$.}
    \label{fig:SI4}
    \end{figure}

We assume that the hBN disk is placed with its flat face parallel to the x-y plane and the air/hBN/PEC layers are arranged along the $z$ axis (as shown in Fig. \ref{fig:SI4}a). We model hBN as a dispersive, uniaxial material with its optic axis parallel to the $z$ direction such that the diagonal elements of the relative permittivity tensor in the crystal plane are $\epsilon_{xx}= \epsilon_{yy}=\epsilon_x(\omega)$ and along the crystal axis is $\epsilon_{zz}=\epsilon_z(\omega)$. We will account for the frequency dependence of $\epsilon_x$ and $\epsilon_z$ in all subsequent calculations but we shall omit their explicit $\omega$ dependence for notational simplicity in the equations that follow. The frequency dependent permittivity components are calculated according to the parameters in Ref. \cite{Caldwell2014}.

We consider modes propagating along the $x$ direction in the layered structure with the hBN layer extending from $z=0$ to $z=d$. Since the HPhPs are TM waves, we focus exclusively on modes where ($E_x,\ H_y,\ E_z$) are the only non-zero electromagnetic field components. We start with Maxwell's curl equations and following some simplification, we obtain
\begin{equation}
    \left[\frac{\partial^2}{\partial z^2} + \epsilon_x\left( k_0^2 - \frac{\beta^2}{\epsilon_z}\right) \right] H_y=0,
    \label{eq:1}
\end{equation}
where $k_0$ is the free space wavevector and $\beta=n_{\mathrm{eff}}k_0$ is the propagation constant along the $x$ direction. The HPhPs are confined in the volume of the hBN and decay in the air. Therefore, we assume that $H_y$ has solutions of the form
\begin{equation}
    \begin{split}
        H_y(0<z<d)&=Ae^{ik_zz} + B e^{-ik_zz}, \\
        H_y(z>d)&=Ce^{-\alpha(z-d)},
    \end{split}
    \label{eq:2}
\end{equation}
where $k_z^2=\epsilon_x\left( k_0^2 - \frac{\beta^2}{\epsilon_z}\right)$, $\alpha=\sqrt{\beta^2-k_0^2}$ and $A,\ B,\ C$ are arbitrary constants. The tangential electric field component at the hBN/PEC interface is zero ($E_x|_{z=0}=0$) and the tangential components $E_x$ and $H_y$ are continuous across the air/hBN interface at $z=d$. Using these boundary conditions, we obtain the dispersion relation for the HPhPs
\begin{equation}
    \tan(k_zd)=\frac{\alpha\epsilon_x}{k_z}.
    \label{eq:3}
\end{equation}
Eq. \ref{eq:3} is a transcendental equation that must be solved to obtain $n_{\mathrm{eff}}$ and may be re-written in a less compact form that reveals the existence of multiple HPhP modes and the dependence of $n_{\mathrm{eff}}$ on the dispersive hBN permittivity as
\begin{equation}
    k_0d\sqrt{\epsilon_x\left( 1-\frac{n_{\mathrm{eff},l}^2}{\epsilon_z} \right)} = \arctan\left( \sqrt{\frac{\epsilon_x\epsilon_z(n_{\mathrm{eff},l}^2-1)}{\epsilon_z-n_{\mathrm{eff},l}^2}} \right) + l\pi,
    \label{eq:4}
\end{equation}
where $l$ is the HPhP mode order for the layered structure and $n_{\mathrm{eff},l}$ is the effective index for the corresponding mode. The dispersion relation for HPhP modes calculated from eq. \ref{eq:4} for different values of $l$ is shown in Fig. \ref{fig:SI4}b. We see that $n_{\mathrm{eff},0}\sim20$ for even the lowest order $l=0$ mode for the frequencies of interest in RSII, demonstrating the deeply sub-wavelength nature of the HPhPs. We find that the HPhPs have finite loss with $\mathrm{Im}(n_{\mathrm{eff}}) / \mathrm{Re}(n_{\mathrm{eff}})<0.15$. Given that decay constant is much lesser than the propagation constant, we neglect $\mathrm{Im}(n_{\mathrm{eff}})$ in the subsequent calculations.

\subsection{Modes of the hBN microdisk}
\label{sec:2D}
\begin{figure}[H]
    \centering
    \includegraphics[width=1\linewidth]{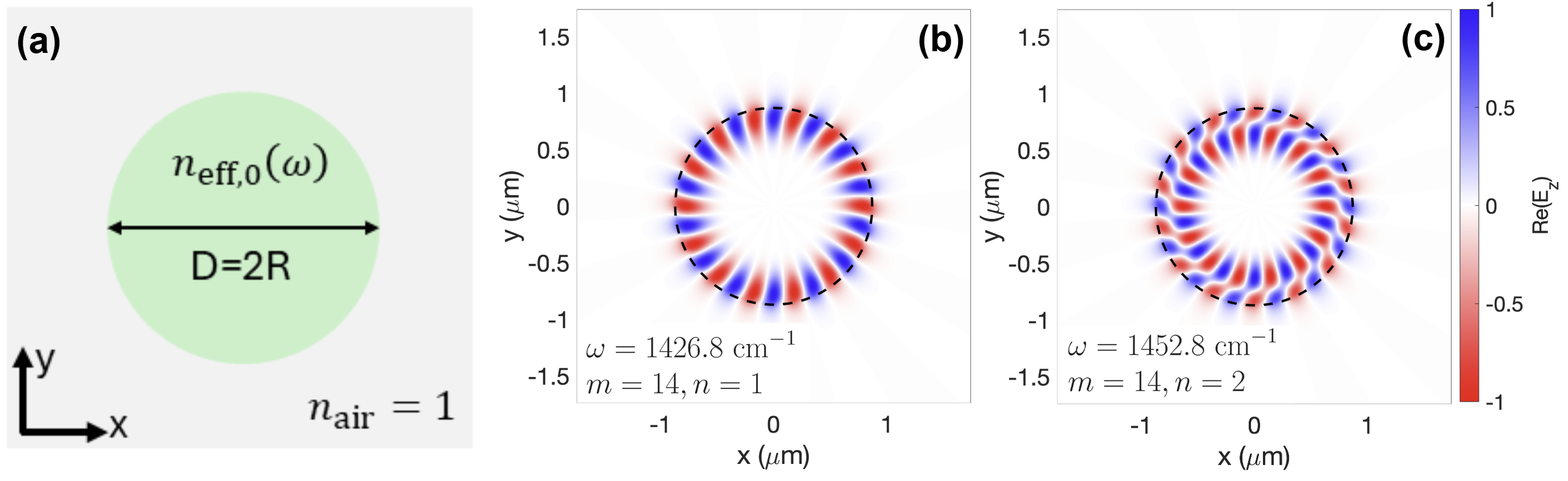}
    \caption{(a) Schematic of the hBN disk with radius $R$ and refractive index $n_{\mathrm{eff},0}$. (b) Spatial $\mathrm{Re}(E_z)$ profile of the $m=14,\ n=1$ mode of a hBN disk with $d=30$ nm and $R=870$ nm at $\omega=1426.8\ \mathrm{cm}^{-1}$. The dashed black line shows the edge of the disk. (c) Same as (b) but for the $m=14,\ n=2$ mode at $\omega=1452.8\ \mathrm{cm}^{-1}$.}
    \label{fig:S5}
\end{figure}

While multiple HPhP modes are supported by the hBN, the $l=0$ mode is dominant in experiments \cite{Ambrosio2018}. Therefore, we will focus exclusively on the $l=0$ mode and neglect higher $l$ modes in our calculation \cite{Tamagnone2018}. The problem of $z$-confinement of the HPhPs has been solved in section \ref{sec:1D}, and the frequency dependent solution $n_{\mathrm{eff},0}$ obtained from eq. \ref{eq:4} encapsulates the effect of the HPhP being confined to the hBN layer along $z$. We use $\mathrm{Re}(n_{\mathrm{eff},0})$ as the refractive index of a 2D disk of radius $R$ in the $z=0$ plane embedded in an air environment ($n_{\mathrm{air}}=1$) as shown in Fig. \ref{fig:S5}a and solve for its eigenmodes.

We choose the cylindrical ($\hat{\rho},\hat{\phi},\hat{z}$) coordinate system in our calculations since it is naturally suited to the geometry at hand. The wave equation in cylindrical coordinates decouples such that the $E_z$ and $H_z$ components independently satisfy the scalar Helmholtz equation $\left[\nabla^2 + \epsilon_rk_0^2\right] \psi=0$, where $\psi=E_z$ or $H_z$. We solve the wave equation for $E_z$ since the HPhP is a TM mode with $E_z\neq0$. Since we are solving for the modes a 2D disk, we assume that $E_z$ is invariant along $z$, i.e., $\frac{\partial}{\partial z}\to ik_z=0$. With these assumptions, the solution to the wave equation has the form 
\begin{equation}
    E_z\propto J_m(k_0\sqrt{\epsilon_r}\rho)e^{im\phi},
    \label{eq:5}
\end{equation}
where $J_m$ is the Bessel function of order $m$, and $m$  must be an integer that  represents the number of full waves around the circumference of the disk \cite{JacksonBook}.

Following eq. \ref{eq:5} we assume that the solution inside the disk is $E_z(\rho<R,\phi)=PJ_m(k_d\rho)e^{im\phi}$, and the solution outside the disk in the air environment is assumed to be $E_z(\rho>R,\phi)=QK_m(k_0\rho)e^{im\phi}$ where $P,Q$ are arbitrary constants, $k_d=k_0n_{\mathrm{eff},0}$ and we have used modified Bessel functions of the second kind ($K_m$) for $\rho>R$ since we are interested in modes that are bound to the hBN disk with evanescent tails in the air environment \cite{Wait1967}. The tangential field components $E_z$ and $H_{\phi}$ must be continuous at the disk-air interface at $r=R$. Applying these boundary conditions and following some algebraic simplification, we obtain the dispersion relation for the modes of the disk
\begin{equation}
    k_d\frac{J'_m(k_dR)}{J_m(k_dR)}=k_0\frac{K'_m(k_0R)}{K_m(k_0R)},
    \label{eq:6}
\end{equation}
where the prime indicates the derivative. We use the recurrence relation for derivatives of the Bessel functions to re-write eq. \ref{eq:6} as
\begin{equation}
    k_d\frac{J_{m+1}(k_dR)}{J_m(k_dR)}=k_0\frac{K_{m+1}(k_0R)}{K_m(k_0R)}.
    \label{eq:7}
\end{equation}
Eq. \ref{eq:7} is a transcendental equation that must be solved in conjunction with eq. \ref{eq:4} to obtain the $\omega-m$ dispersion relation for the HPhP modes of the hBN disk. For a given value of $m$, multiple solutions to eq. \ref{eq:7} may be found at different values of $\omega$. These solutions correspond to eigenmodes of the hBN disk with different counts of the number of anti-nodes in the radial direction. We denote the number of anti-nodes in the radial direction with $n$ and the lowest frequency solution corresponds to the $n=1$ eigenmode. The spatial $\mathrm{Re}(E_z)$ profile of the $n=1$ and $n=2$ modes for $m=14$ for a hBN disk with $d=30$ nm and $R=870$ nm is shown in Fig. \ref{fig:S5}b and \ref{fig:S5}c, respectively.
\newpage

\newpage
\section{Comparison of remote excitation schemes for hyperbolic phonon-polaritons}

Several approaches have been developed to excite hyperbolic phonon-polaritons without relying on direct tip launching in s-SNOM experiments. These include metallic scatterers, buried metallic structures, wrinkles, dielectric gratings, and various resonant launcher geometries. Since the present work employs an auxiliary cavity as a remote launcher, it is useful to place this approach within the broader context of existing excitation schemes.

A direct quantitative comparison of excitation efficiencies reported in different studies is, however, not straightforward. The coupling efficiency depends strongly on sample-specific parameters, including the hBN thickness, substrate material, dielectric environment, excitation wavelength, and polariton propagation length. Since these parameters vary substantially between studies, direct normalization of reported excitation efficiencies is generally not meaningful.

Instead, a qualitative comparison based on the underlying excitation mechanism and suitability for exciting confined resonant modes provides a more physically relevant perspective. Table~S1 summarizes several representative remote-excitation strategies reported for hBN phonon-polaritons.

\begin{table*}[htbp]
\centering
\small
\renewcommand{\arraystretch}{1.5}
\setlength{\tabcolsep}{4pt}

\begin{tabular}{|p{3.0cm}|p{2.8cm}|p{1.8cm}|p{3.8cm}|p{3.2cm}|p{1.3cm}|}
\hline
\textbf{Method} &
\textbf{Nature} &
\textbf{Separation} &
\textbf{Perturbation} &
\textbf{Tech. availability} &
\textbf{Refs.} \\
\hline

Metal scatterer on hBN &
Non-resonant &
No &
Strong (near-field scattering at interface) &
Standard photolithography/EBL &
Ref.~\cite{ni2021long} \\
\hline

Metallic grating couplers &
Momentum-matched plasmonic / polaritonic &
No &
Strong (periodic field modulation in-plane) &
Standard photolithography/EBL &
Ref.~\cite{ivchenko2024phonon} \\
\hline

Buried metal structures under hBN &
Non-resonant &
Yes &
Weak (boundary-defined evanescent field injection, no resonant mode) &
Advanced CMOS-compatible nanofab &
Ref.~\cite{duan2017launching} \\
\hline

hBN wrinkles/folds &
Non-resonant &
No &
Weak (smooth dielectric modulation) &
Uncontrolled / natural &
Ref.~\cite{duan2017launching} \\
\hline

hBN flake edges &
Non-resonant &
No &
Moderate (sharp dielectric discontinuity) &
Intrinsic &
Ref.~\cite{dai2017efficiency} \\
\hline

hBN cavities at metal/dielectric interface &
Resonant field enhancement &
Yes &
Weak (separating interface, but broken azimuthal symmetry) &
Standard photolithography/EBL &
Ref.~\cite{borodin2026cavity} \\
\hline

Dielectric gratings &
Momentum engineering &
No &
Weak (small index modulation) &
Standard photolithography/EBL &
Ref.~\cite{zhang2022unidirectionally} \\
\hline

\end{tabular}

\caption{Non-tip phonon-polariton launching mechanisms in hBN-based systems classified by excitation physics, spatial separation, and perturbation strength.}
\label{tab:php_launchers}
\end{table*}

The comparison shows that several remote-launching strategies have been successfully employed to excite propagating hyperbolic phonon-polaritons. Rather than claiming superiority over these approaches, the objective of the present work is to demonstrate a launcher specifically suited for exciting whispering-gallery modes in a separate resonator.

Among the existing techniques, buried metallic structures and the auxiliary cavities employed here are conceptually the most closely related, as both utilize an impedance mismatch to generate a spatially separated excitation source while only weakly perturbing the resonator itself. The principal distinction is that the auxiliary cavity supports a localized resonant mode, providing resonant field enhancement before launching the phonon-polaritons into the disk resonator.

As demonstrated in our previous work on cavity-driven phonon-polariton launchers (Borodin \textit{et al.}, Commun. Phys. 2026), the auxiliary cavity produces a substantially stronger launched field than conventional tip excitation while also allowing the emitted wavefront to be engineered through the cavity geometry. Furthermore, the additional simulations presented in this Supplementary Information demonstrate that even a localized defect can excite whispering-gallery modes, but with substantially lower efficiency than the resonant auxiliary cavity. This indicates that the principal advantage of the proposed launcher originates from its resonant field enhancement rather than simply from the presence of a scattering center.

Under realistic experimental conditions, where material losses, finite resonator quality factors, fabrication imperfections, and competing tip-launched signals all reduce the visibility of the modal field distribution, this resonant enhancement can make the difference between readily observable whispering-gallery modes and resonances that remain nearly hidden in the measured near-field response.

\section{On the scattering bottleneck in etching-defined cavities}
\vspace{-1em}
An additional question we aimed to address in this work is identifying the actual bottleneck in etching-defined cavities. To this end, we analyzed the temperature dependence of Q-factors of hBN disks fabricated from natural-abundance hBN and isotopically pure h$^{11}$B$^{14}$N.

The total losses in the system can be expressed as:
\vspace{-1em}
\begin{equation}
\frac{1}{Q_{\rm tot}} = \frac{1}{Q_{\rm rad}} + \frac{1}{Q_{\rm mat}(T)} + \frac{1}{Q_{\rm scat}},
\end{equation}

where $1/Q_{\rm tot}$, $1/Q_{\rm rad}$, $1/Q_{\rm mat}(T)$, and $1/Q_{\rm scat}$ are the total, radiative, material, and scattering quality factors, respectively.

Only $1/Q_{\rm mat}(T)$ depends on temperature, because temperature affects the thermally activated phonon population in the material: lower temperatures lead to reduced phonon–phonon scattering in hBN. By comparing natural-abundance hBN with isotopically pure h$^{11}$B$^{14}$N—where $1/Q_{\rm mat}$ has been shown to be lower\cite{giles2018ultralow}—we aim to observe this effect and decouple the individual contributions to $1/Q_{\rm tot}$. We first examined the temperature dependence of the quality factor for disks with diameters of 1~$\mu$m and 1.5~$\mu$m and a thickness of approximately 30~nm, fabricated from both natural-abundance hBN and h$^{11}$B$^{14}$N. We found that their quality factors are nearly identical within experimental uncertainty and exhibit only a weak temperature dependence (see Section~A).

\vspace{-1em}
\subsection{Nat. Abundance hBN vs h$^{11}$B$^{14}$N}
\begin{figure}[H]
    \centering
    \includegraphics[scale=0.9]{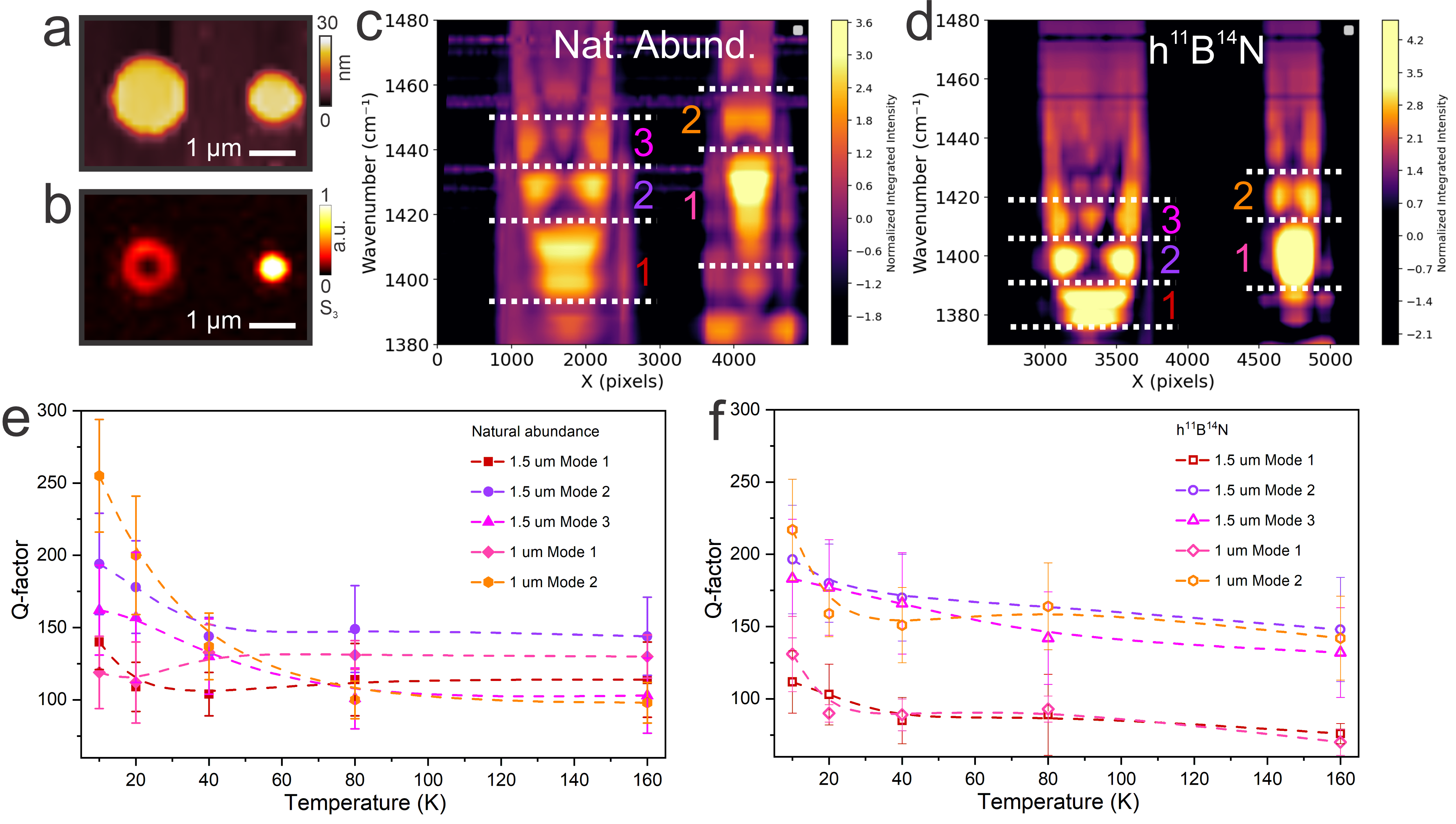}
        \vspace{-1.5em}
    \caption{
    \textbf{Temperature dependence of Q-factors for natural-abundance hBN vs h$^{11}$B$^{14}$N disks.}
    \textbf{a)} AFM topography of 1-$\mu$m- and 1.5-$\mu$m-diameter hBN disks.
    \textbf{b)} s-SNOM amplitude map ($S_3$) at 1430 cm$^{-1}$.
    \textbf{c)} Contour plot composed from linescans recorded in the range $\omega = 1380$–1480 cm$^{-1}$ through the natural-abundance hBN disks at 10 K.
    \textbf{d)} Contour plot composed from linescans recorded in the range $\omega = 1380$–1480 cm$^{-1}$ through the h$^{11}$B$^{14}$N disks at 10 K.
    \textbf{e)} Q-factor vs. temperature for the natural-abundance hBN disk.
    \textbf{f)} Q-factor vs. temperature for the h$^{11}$B$^{14}$N disk.
}

    \label{fig:S6}  
\end{figure}

\vspace{-1em}
Figure~\ref{fig:S6} demonstrates the results of a comparison between natural-abundance hBN disks and h$^{11}$B$^{14}$N disks. As shown in Figures ~\ref{fig:S6}c and ~\ref{fig:S6}d, the modes in the disks are similar but spectrally shifted. This is due to the fact that the TO phonon modes in natural-abundance hBN and h$^{11}$B$^{14}$N are located at different frequencies~\cite{janzen2024boron}.
Figures ~\ref{fig:S6}e and ~\ref{fig:S6}f show Q-factor analysis of the first three modes in 1.5-$\mu$m-diameter hBN disks and the first two modes in 1.0-$\mu$m-diameter hBN disks. As can be seen, the Q-factors and their temperature dependence are quite similar (within experimental error). Moreover, the dependence of Q-factor on temperature is relatively weak. We attribute the similar Q-factors and their weak temperature dependence to the dominant contribution of $1/Q_{\rm scat}$. 
\vspace{-1em}
\vspace{-1em}
To test this assumption, we fabricated an additional set of thicker (54 nm) cavities from h$^{11}$B$^{14}$N. The corresponding results are presented in Section~B.

\subsection{Temperature Dependence of HPhP Q-Factor in a 54-nm-Thick h$^{11}$B$^{14}$N Disk}

\begin{figure}[H]
    \centering
    \vspace{-1.5em}
    \includegraphics[scale=0.8]{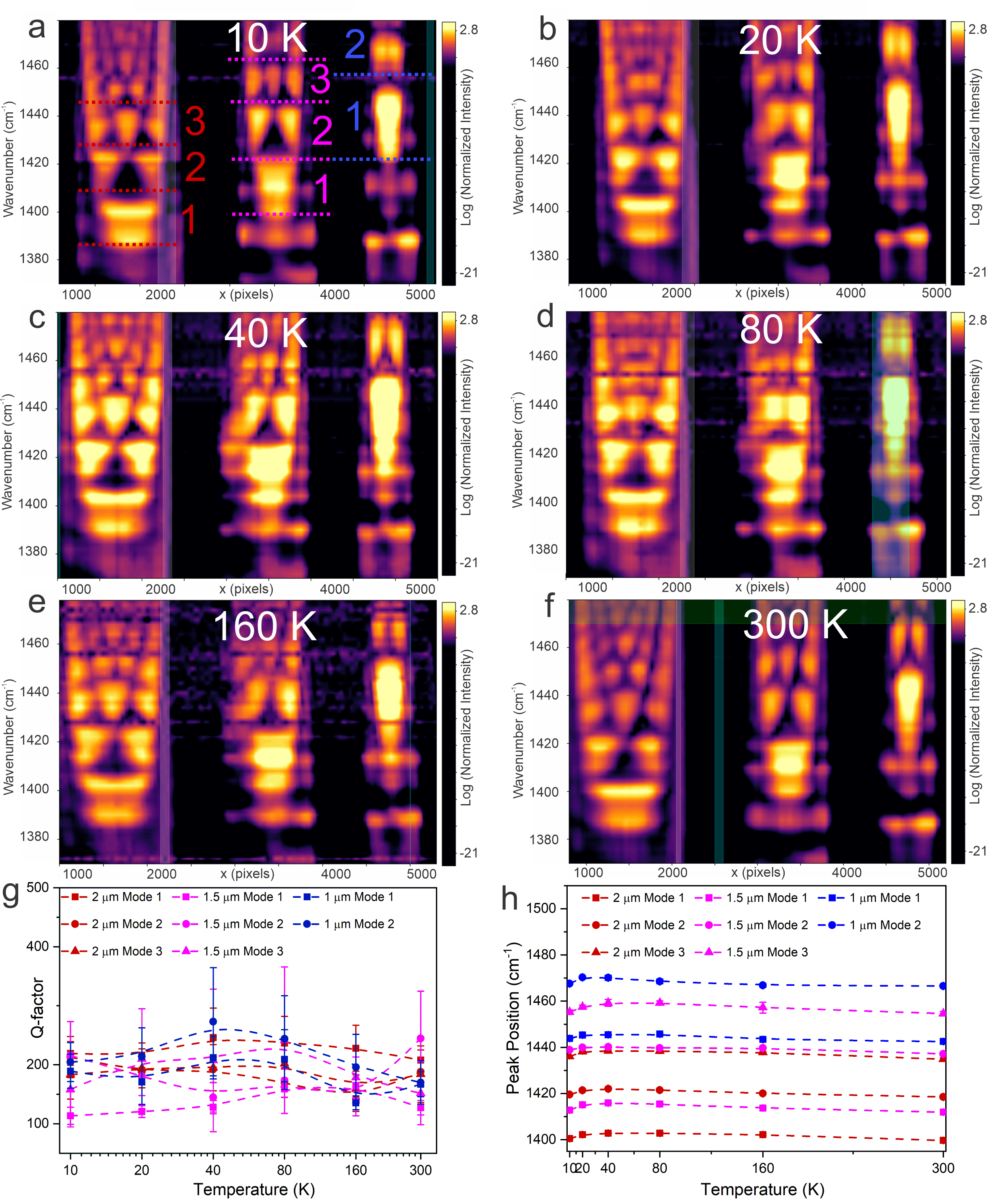}
    \vspace{-1.5em}
   \caption{\textbf{Temperature dependence of Q factors for 54-nm-thick h$^{11}$B$^{14}$N disks.}
\textbf{a--f)} Contour plot composed of line scans recorded in the range $\omega = 1370$--1480~cm$^{-1}$ through the h$^{11}$B$^{14}$N disks from 10 to 300~K.
\textbf{g)} Temperature dependence of the HPhP Q factor for the main modes in 2-$\mu$m, 1.5-$\mu$m, and 1-$\mu$m disks.
\textbf{h)} Temperature dependence of the peak position for the main modes in 2-$\mu$m, 1.5-$\mu$m, and 1-$\mu$m disks.
}
\vspace{-1em}
    \label{fig:S7}  
\end{figure}

Figure~\ref{fig:S7}a--f demonstrate the temperature dependence of round-trip modes localized in 54-nm-thick h$^{11}$B$^{14}$N disks in the spectral range from 1370 to 1480~cm$^{-1}$. Figure~\ref{fig:S7}g summarizes the temperature dependence of the Q factors for 2-$\mu$m, 1.5-$\mu$m, and 1-$\mu$m disks. As can be seen, the Q factor does not depend on temperature within the experimental error. This observation is consistent with the scattering bottleneck discussed in Section~IVA, ``On the scattering bottleneck in etching-defined
cavities'' Figure~\ref{fig:S7}h shows the temperature dependence of the mode peak positions, which exhibit a maximum at approximately 20~K.

In this case, the quality factors remain nearly constant over a wide temperature range from 10 to 300~K.

Assuming that only $1/Q_{\mathrm{mat}}$ exhibits a significant temperature dependence, these observations suggest that, in etching-defined cavities, scattering from defective edges ($1/Q_{\mathrm{scat}}$) constitutes the dominant bottleneck that largely determines $1/Q_{\mathrm{tot}}$. For thicker cavities, a larger number of defects participate in scattering, rendering $Q_{\mathrm{tot}}$ effectively temperature independent, consistent with our observations for 54~nm-thick disks.

While these results are not yet conclusive and require further investigation, they provide important insight. On the one hand, dielectric-contrast-defined cavities may offer a more promising route toward achieving high-$Q$ polaritonic resonators. On the other hand, for thick etching-defined hBN cavities, phonon–phonon scattering appears to play a negligible role in determining $1/Q_{\mathrm{tot}}$, whereas edge-related scattering represents the true bottleneck for further quality-factor enhancement. We suggest that systematic studies of the etching process, including plasma composition, etching rate, and edge roughness, may shed further light on this issue and enable optimization strategies to minimize edge-related scattering losses.

\bibliographystyle{ieeetr} 
\bibliography{references_SI.bib}